\documentclass[twoside]{article}

\usepackage[accepted]{stylefile}

\usepackage[round]{natbib}

\usepackage{graphicx}
\usepackage{footnote}
\usepackage[colorlinks,linkcolor=black,citecolor=black,urlcolor=blue,filecolor=blue,backref=false]{hyperref}
\usepackage{microtype}
\usepackage{amssymb}

\usepackage{blkarray}

\usepackage{mathtools}

\usepackage{enumerate}
\usepackage{enumitem}

\usepackage{booktabs}

\newenvironment{nalign*}{
    \begin{equation*}
    \begin{aligned}
}{
    \end{aligned}
    \end{equation*}
    \ignorespacesafterend
}

\newenvironment{nalign}{
    \begin{equation}
    \begin{aligned}
}{
    \end{aligned}
    \end{equation}
    \ignorespacesafterend
}

\begin{document}

\twocolumn[

\aistatstitle{Uncertainty-aware Sensitivity Analysis Using R\'{e}nyi Divergences}

\aistatsauthor{ Topi Paananen$^1$ \And  Michael Riis Andersen$^2$ \And Aki Vehtari$^1$ }

\aistatsaddress{ {\tt topi.paananen@aalto.fi} \And {\tt miri@dtu.dk} \And {\tt aki.vehtari@aalto.fi} } 
\vspace{-1.6em}
\aistatsaddress{ $^1$Helsinki Institute for Information Technology, HIIT \\ $^1$Aalto University, Department of Computer Science \\
$^2$Department of Applied Mathematics and Computer Science, Technical University of Denmark }

\runningauthor{Topi Paananen, Michael Riis Andersen, Aki Vehtari}

]

\begin{abstract}
For nonlinear supervised learning models, assessing the importance
of predictor variables or their interactions
is not straightforward because it
can vary in the domain of the variables.
Importance can be assessed locally with sensitivity analysis using general methods that
rely on the model's predictions or their derivatives.
In this work, we extend derivative based sensitivity analysis to a Bayesian setting
by differentiating the R\'{e}nyi divergence of
a model's predictive distribution.
By utilising the predictive distribution instead of a point prediction, the model uncertainty
is taken into account in a principled way.
Our empirical results on simulated and real data sets
demonstrate accurate and reliable identification of important variables and interaction effects compared to
alternative methods.
\end{abstract}

\section{Introduction} \label{sec:intro}

\begin{figure}[h]
  \centering
    \includegraphics[width=0.48\textwidth]{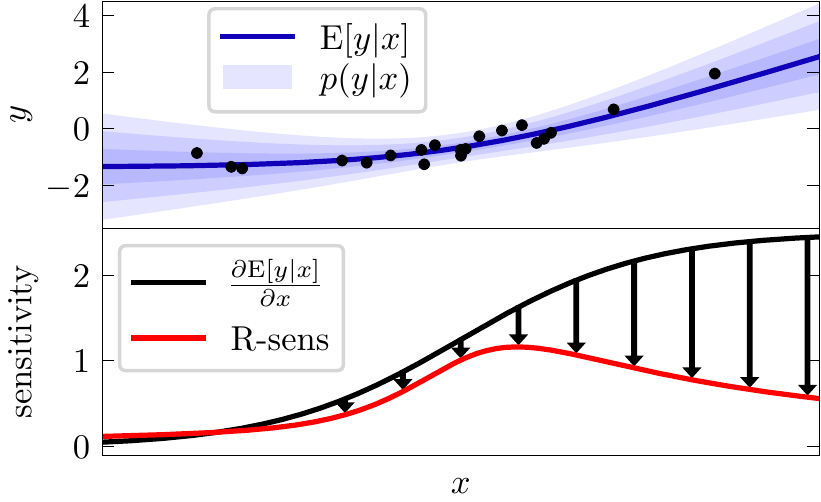}
\caption{Top: Example of data and a probabilistic model with a Gaussian predictive distribution $p(y | x)$. The different shades of blue represent 1, 2, and 3 standard deviations of the predictive distribution. Bottom:
The derivative of $\text{E}[y | x]$ with respect to $x$ (black) represents the naive
sensitivity of the model's predictions to changes in $x$.
The R-sens method proposed in this work (red) represents
uncertainty-aware sensitivity as given by
differentiating a R\'{e}nyi divergence of predictive distributions, which
adjusts the sensitivity according to uncertainty about $y$.}
\label{fig:illustr2}
\end{figure}

Identifying important features and interactions
from complex data sets and models remains a topic of active research.
This is a fundamental problem with important applications in many scientific
disciplines.
Often the goal is to improve understanding of the model, but
the identified features and interactions can also be used
to build a simpler or more interpretable surrogate model.

For models that can capture
nonlinear effects and interactions, the typical approach is to
assess the contributions of individual predictors or interactions on the model's prediction at an individual observation.
One approach is sensitivity analysis, which
evaluates the change in predictions to
small perturbations in the predictor values.
For example, the
partial derivative of
the model's prediction with respect to the predictors can be a measure of importance~\citep{cacuci2003sensitivity,oakley2004probabilistic,cacuci2005sensitivity,guyon2003introduction}.
Since the derivative can
vary from positive to negative in the domain of the predictors, most approaches use absolute or squared
derivatives averaged from the observations~\citep{ruck1990feature,dorizzi1996variable,czernichow1996architecture,refenes1999neural,leray1999feature,sundararajan2017axiomatic,cui2019recovering}.
A similar approach is popular in image classification, where derivatives with respect to
each pixel are called saliency maps~\citep{simonyan2013deep,zeiler2014visualizing,guidotti2018survey}.
The average predictive comparison of~\citet{gelman2007average}
uses the difference quotient of two predictions without taking the limit.
By extending to cross-derivatives with respect to two predictors, one can also measure
the interaction effect of predictors~\citep{friedman2008predictive,cui2019recovering}.
A closely related approach for sensitivity analysis is to directly estimate the contribution of predictor main effects or interactions
to the variance of the target variable~\citep{im1993sensitivity,homma1996importance,oakley2004probabilistic,saltelli2002sensitivity}.

Recently, approaches that evaluate the differences of predictions in permuted training observations
have gained popularity in machine learning.
For example,~\citet{fisher2019all} permute the observations of a single predictor, and
examine the loss in predictive ability compared to the original data.
Shapley values use permutations to assess the
average marginal contribution of a predictor to a specific observation~\citep{shapley1953value,vstrumbelj2014explaining}.
\citet{lundberg2018consistent} extended this approach to
evaluate second-order interactions
based on the Shapley interaction index~\citep{fujimoto2006axiomatic}.
\citet{friedman2008predictive} and \citet{greenwell2018simple} use permutations and
partial dependence functions~\citep{friedman2001greedy} to construct
statistics that measure the strength of pairwise interactions.
The individual conditional expectation plots of~\citet{goldstein2015peeking} can also be used
to identify interactions, but they rely
on visualisation only.

The contributions of this work are summarised as follows.
First, we present a novel method that generalises derivative and Hessian based sensitivity analysis
to a Bayesian setting
for models with a parametric predictive distribution or its approximation.
Instead of using the first or second derivatives of the mean prediction of the model,
we instead differentiate the R\'{e}nyi divergence from
one predictive distribution to another, which takes into account the epistemic uncertainty
of the predictions.
Figure~\ref{fig:illustr2} gives an illustration of this method.
Second, we show that our method is an analytical generalisation and extension of a previous
finite difference method~\citep{paananen2019variable}.
Third, we show empirically that our proposed method can improve the accuracy of
sensitivity analysis
in situations where
the used model has significant predictive uncertainty.
Code for our method and simulations is available at~\href{https://github.com/topipa/rsens-paper}{https://github.com/topipa/rsens-paper}.

\section{Uncertainty-Aware Sensitivity} \label{sec:method}

Consider a supervised learning
model trained on data $( \mathbf{X}, \mathbf{y} )$, where $\mathbf{X} \in \mathbb{R}^{N \times D}$ is the design matrix
and $\mathbf{y} \in \mathbb{R}^{N}$ is the vector of target observations.
Let us denote the prediction function of the model for the target variable $y$ as
$f  (\mathbf{x}^{*})$.
Derivative based sensitivity analysis can be used to assess
the local sensitivity of $f$ to the different predictors $(x_d)_{d = 1}^D$.
The sensitivity can be quantified by the partial derivative
\begin{nalign*}
\frac{\partial f  (\mathbf{x}^{*})}{\partial x_d^{*}}    .
\end{nalign*}
Absolute values of local derivatives can be aggregated over the empirical distribution of $\mathbf{x}$
to obtain a global importance estimate for $x_d$, the expected absolute derivative (EAD)
\begin{nalign*}
\text{EAD} (x_d) = \text{E}_{p (\mathbf{x})} \left [\left | \frac{\partial f  (\mathbf{x})}{\partial x_d} \right | \right]   .
\end{nalign*}
Similarly, absolute values of the
elements of the Hessian matrix of $f$, that is, the
second derivatives with respect to $x_d$ and $x_e$, quantify the sensitivity
to the joint interaction effect of $x_d$ and $x_e$.

In this section, we present our proposed R-sens method that extends derivative and Hessian based
sensitivity analysis
methods to a Bayesian setting where the evaluated model not only has a function
for point predictions, but a \emph{predictive distribution}
$p  (y^{*}) $.
For now, we only consider predictive distributions that have some
parametric form, which can be obtained exactly in closed form or it can be an approximation.
Because the predictive distribution is obtained by integrating over
posterior uncertainty for the model parameters,
it is important to utilise this uncertainty
in sensitivity analysis as well.

To formulate a derivative based sensitivity measure for a model with a predictive distribution,
we need a suitable functional of the predictive distribution, which to differentiate.
We choose a family of statistical divergences called
R\'{e}nyi divergences due to their convenient properties, which we discuss later in this section.
R\'{e}nyi divergence of order $\alpha$ is defined for two probability mass functions $P = (p_1 , ..., p_n)$ and $Q = (q_1 , ..., q_n)$ as
\begin{nalign*}
\mathcal{D}_{\alpha} [P || Q] = \frac{1}{\alpha - 1} \log \left ( \sum_{i = 1}^{n} \frac{p_i^{\alpha}}{q_i^{\alpha - 1}} \right )
\end{nalign*}
when $0 < \alpha < 1$ or $1 < \alpha < \infty$~\citep{renyi1961measures,van2014renyi}.
The definition generalises to continuous spaces by replacing the probabilities by densities and the sum by an integral.
The divergences for values $\alpha = 0, 1$, and $\infty$ are obtained as limits.
The most well-known R\'{e}nyi divergence
is the Kullback-Leibler divergence which is obtained in the limit $\alpha \rightarrow 1$~\citep{kullback1951information}.

Let us consider a model with a predictive distribution
parametrised by a vector $\boldsymbol{\lambda}^{*} = ( \lambda_1^{*} , ... , \lambda_{M}^{*} )$, which possibly depends on
$\mathbf{x}^{*}$.
Let us denote the predictive distribution for $y$ conditional on predictor values $\mathbf{x}^{*}$ as
\begin{nalign*}
p  (y^{*})  \equiv p  (y^{*} | \boldsymbol{\lambda}^{*} (\mathbf{x}^{*})) .
\end{nalign*}
Keeping $\mathbf{x}^{*}$ fixed, we denote the R\'{e}nyi divergence of order $\alpha$ between
two predictive distributions as a function of $\mathbf{x}^{**}$ as
\begin{nalign*}
\mathcal{D}_{\alpha}^p [\mathbf{x}^{**}] \equiv \mathcal{D}_{\alpha} [ \, p  (y^{*} | \boldsymbol{\lambda}^{*}(\mathbf{x}^{*}) ) || p  (y^{*} | \boldsymbol{\lambda}^{*}(\mathbf{x}^{**} )) ].
\end{nalign*}
We formalise
the sensitivity of the predictive distribution to a change in a single predictor variable
by differentiating the R\'{e}nyi divergence
in the limit
when the distributions coincide, that is when $\mathbf{x}^{**} = \mathbf{x}^{*}$.
However, because
R\'{e}nyi divergences obtain their minimum value when the two distributions
coincide, the first derivative at this point is always zero.
Hence, we formulate the uncertainty-aware sensitivity measure 
with respect to
the predictor $x_d$ using the second derivative
\begin{align} \label{eq:2der1}
&  \frac{\partial^2 \mathcal{D}_{\alpha}^p [\mathbf{x}^{**}]}{(\partial x_d^{**})^2} \bigg|_{ \mathbf{x}^{**} =  \mathbf{x}^{*}} = \\
 &  \bigg(\frac{\partial \boldsymbol{\lambda}^{*} (\mathbf{x}^{*})}{\partial x_d^{*}} \bigg)^{T}   \mkern-8mu
\mathbf{H}_{\boldsymbol{\lambda}^{*} (\mathbf{x}^{**})} ( \mathcal{D}_{\alpha}^p [\mathbf{x}^{**}] )
 \bigg(\frac{\partial \boldsymbol{\lambda}^{*} (\mathbf{x}^{**})}{\partial x_d^{**}} \bigg) \mkern-4mu \bigg|_{ \mathbf{x}^{**} =  \mathbf{x}^{*}} , \nonumber
\end{align}
where
$\mathbf{H}_{\boldsymbol{\lambda}^{*} (\mathbf{x}^{**})}$ is the Hessian matrix of the R\'{e}nyi divergence with
second order derivatives with respect to $\boldsymbol{\lambda}^{*} (\mathbf{x}^{**})$.

The sensitivity measure in equation~(\ref{eq:2der1}) has two kinds of partial derivatives: (i) second order derivatives
of the R\'{e}nyi divergence with respect to the parameters $\boldsymbol{\lambda}^{*}$ of the predictive
distribution, and (ii) first order derivatives of the parameters $\boldsymbol{\lambda}^{*}$ with respect to the predictor $x_d^{*}$.
These are obtained as follows:
\begin{enumerate}
\item[(i)] For sufficiently regular parametrisations, the second order
Taylor approximation of the Kullback-Leibler divergence ($\alpha = 1$)
gives an approximate equivalence between the Hessian of the divergence and
the Fisher information matrix of $p  (y^{*})$
in the limit $\mathbf{x}^{**} - \mathbf{x}^{*} \rightarrow 0$~\citep{kullback1959statistics,van2014renyi}.
\citet{haussler1997mutual} state that this generalises to any
R\'{e}nyi divergence with $0 < \alpha < \infty$, leading to the relation
\begin{nalign} \label{eq:fisheridentity}
\mathbf{H}_{\boldsymbol{\lambda}^{*} (\mathbf{x}^{**})} ( \mathcal{D}_{\alpha}^p [\mathbf{x}^{**}] ) |_{ \mathbf{x}^{**} =  \mathbf{x}^{*}} \approx \alpha
\mathcal{I}(\boldsymbol{\lambda}^{*} (\mathbf{x}^{*})) ,
\end{nalign}
where $\mathcal{I}(\boldsymbol{\lambda}^{*}(\mathbf{x}^{*}))$ is the Fisher information matrix of
the distribution $p  (y^{*} | \boldsymbol{\lambda}^{*}(\mathbf{x}^{*}) )$.
\item[(ii)] The partial derivative of the parameter $\lambda^{*}_k$ with respect to predictor variable $x^{*}_d$ depends
on the model where the predictive distribution is from.
\end{enumerate}

We define R-sens, an uncertainty-aware sensitivity measure for predictor $x_d$
at $\mathbf{x}^{*}$ as
\begin{nalign} \label{eq:kldiff}
\hspace{-0.3cm} \text{R-sens} (\mathbf{x}^{*},x_d, \alpha) \equiv \sqrt{ \alpha  \bigg(\frac{\partial \boldsymbol{\lambda}^{*}}{\partial x_d^{*}} \bigg)^{T}   \hspace{-0.1cm}
\mathcal{I}(\boldsymbol{\lambda}^{*} (\mathbf{x}^{*}))
\bigg(\frac{\partial \boldsymbol{\lambda}^{*}}{\partial x_d^{*}} \bigg)  } .
\end{nalign}
When using R-sens in sensitivity analysis for comparing predictors, the value of $\alpha$ does not matter
in practice
and we use the value $\alpha = 1$.

In a similar fashion as above, we
generalise the Hessian based sensitivity to a Bayesian predictive distribution
by differentiating the R\'{e}nyi divergence four times, i.e. twice with respect to two predictors.
The full fourth derivative contains cross-derivative terms, which we drop for two reasons.
First, based on our experiments we concluded that
the simplified formula we use is better at identifying interactions, meaning that
the dropped terms do not contain useful information about the interaction effect between
$x_d$ and $x_e$.
Second, the
simplified formula is similar to the R-sens measure and is thus more easily interpretable
and computationally cheaper.
We define R-sens$_2$, the uncertainty-aware sensitivity measure for the interaction effect between variables
$x_d$ and $x_e$ as
\begin{nalign} \label{eq:kldiff2}
& \, \text{R-sens}_2 \, (\mathbf{x}^{*},(x_d,x_e), \alpha) \\
 \equiv  & \, \sqrt{ \alpha  \left (\frac{\partial^2 \boldsymbol{\lambda}^{*}}{\partial x_d^{*} \partial x_e^{*}} \right )^{T} 
\mathcal{I}(\boldsymbol{\lambda}^{*} (\mathbf{x}^{*}))
\left (\frac{\partial^2 \boldsymbol{\lambda}^{*}}{\partial x_d^{*} \partial x_e^{*}} \right )} .
\end{nalign}
In the supplementary material, we show the full equation and an illustration of the benefit
of equation~(\ref{eq:kldiff2}) compared to the full fourth derivative.

\subsection{Location-Scale Family}

For distributions in the location-scale family ($p(y | \lambda_1, \lambda_2) = g((y-\lambda_1)/\lambda_2)/\lambda_2$), the Fisher information of the location parameter is independent of the location parameter~\citep[Ex. 20]{shao2006mathematical}.
This has two implications. First, if the predictive distribution is in the location-scale family,
the R-sens and R-sens$_2$ measures
are direct extensions
to the absolute derivative or absolute Hessian of the mean prediction.
The extension is twofold, as they introduce the derivative of the scale parameter and possible auxiliary parameters,
and also multiplication with the Fisher information matrix.
The uncertainty-aware measures can be also viewed as the Mahalanobis norm
of the differentiated parameters of the predictive distribution instead of a simple Euclidean norm.
Second, for a sufficiently regular model, where the posterior uncertainty vanishes in the asymptotic regime due to
the Bernstein-von Mises theorem~\citep{walker1969asymptotic} such that 
the predictive distribution converges to a limiting distribution in the location-scale family,
R-sens tends to the absolute derivative of the
mean prediction.
Here, we have to assume that the Fisher information exists and converges to
the Fisher information of the limiting distribution.

\subsection{Illustrative Example} \label{sec:illustr_lm}

To illustrate the effects of the different components in equation~(\ref{eq:kldiff}),
we analyse a Bayesian linear regression model as an example.
We use the standard Gaussian likelihood with noise variance $\sigma^2$
and denote the regression coefficients as $\boldsymbol{\beta}$.
Using an improper uniform prior on $(\boldsymbol{\beta}, \text{log} \, \sigma)$, integrating over the uncertainty about all the parameters makes the
posterior predictive distribution at any point $\mathbf{x}^{*}$ (treated as a row vector)
\begin{nalign*}
p(y^{*} | \mathbf{X}, \mathbf{y}) & = \text{Student-}t (\text{E} [y^{*}],\text{Var} [y^{*}], \nu), \, \mathrm{where} \\
\text{E} [y^{*}] & = \mathbf{x}^{*} \boldsymbol{\widehat{\beta}} ,\\
\text{Var} [y^{*}] & = s^2 (1 + \mathbf{x}^{*} (\mathbf{X}^{T} \mathbf{X})^{-1} \mathbf{x}^{* T}), \\
\nu & = N - D ,\\
\boldsymbol{\widehat{\beta}} & = (\mathbf{X}^{T} \mathbf{X})^{-1} \mathbf{X}^{T} \mathbf{y} ,\\
s^2 & = \frac{ (\mathbf{y} - \mathbf{X} \boldsymbol{\widehat{\beta}})^{T} (\mathbf{y} - \mathbf{X} \boldsymbol{\widehat{\beta}}) }{N - D} .
\end{nalign*}
Here, $\nu$ represents the
degrees of freedom, $N$ and $D$ are the number of observations and predictor variables, respectively, and $\boldsymbol{\widehat{\beta}}$ are the maximum likelihood estimates
of the regression coefficients.
The derivative of $\nu$ with respect to
$x^{*}_d$ is zero, and the derivatives of the other two parameters of $p(y^{*} | \mathbf{X}, \mathbf{y})$ are
\begin{nalign*}
\frac{\partial \text{E} [y^{*}]}{\partial x^{*}_d} & = \widehat{\beta}_d ,\\
\frac{\partial \text{Var} [y^{*}]}{\partial x^{*}_d} & = 2 s^2  [ (\mathbf{X}^{T} \mathbf{X})^{-1} \mathbf{x}^{* T}]_d \equiv 2 s^2 V_d .
\end{nalign*}

Multiplying these with the Fisher information matrix of the Student-$t$ predictive distribution,
the R-sens sensitivity measure for the predictor $x_d$ from equation~(\ref{eq:kldiff}) evaluates to
\begin{nalign} \label{eq:2der_lm}
\text{R-sens} (\mathbf{x}^{*}, x_d, \alpha = 1) =
\sqrt{ \frac{ (\nu+1) \widehat{\beta}_{\, d}^{\, 2}  + \frac{ + 2 \nu s^4  V_d^2 }{\text{Var} [y^{*}]}  }{(\nu+3) \text{Var} [y^{*}]} }
 .
\end{nalign}
The two summands have the following interpretations:
In the absence of the second term, the measure
would be proportional to $|\widehat{\beta}_d|$ divided by the
standard deviation of the predictive distribution.
The first term thus measures the absolute derivative of the mean prediction, but
predictions with high uncertainty are given less weight.
Also the second term in equation~(\ref{eq:2der_lm}) quantifies the amount of uncertainty in the
predictive distribution, but in a different way. Even if $\widehat{\beta}_d$
would be exactly zero, the second summand is nonzero as long as there is
uncertainty about the model parameters that causes the predictive uncertainty to
vary with respect to $\mathbf{x}^{*}$.
There are thus two separate mechanisms that include
epistemic uncertainty in the sensitivity analysis~\citep{o2004dicing,kendall2017uncertainties}.
As $N$ (and hence also $\nu$) approaches infinity and the posterior
uncertainty vanishes, the R-sens measures for both variables approach
a constant proportional to $|\widehat{\beta}_d| $.

\begin{figure}[h]
  \centering
    \includegraphics[width=0.48\textwidth]{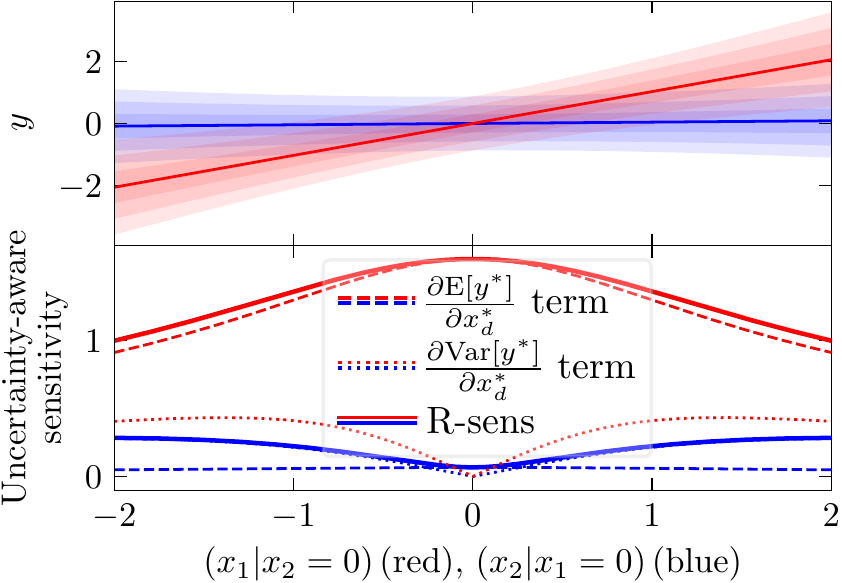}
\caption{Top: Predictive distributions $p(y | x_1, x_2 = 0)$ (red) and $p(y | x_2, x_1 = 0)$ (blue) for the linear regression model in Section~\ref{sec:illustr_lm}.
Bottom: R-sens uncertainty-aware sensitivity measure for
$x_1$ (red) and $x_2$ (blue).
The dashed and dotted lines show the contributions of the two summands in equation~(\ref{eq:2der_lm}).}\label{fig:lm_illustr}
\end{figure}

To visualise the effects of the two terms in equation~(\ref{eq:2der_lm}), we simulated 10 observations from a linear model with
two predictor variables $x_1$ and $x_2$ whose true regression coefficients are $\beta_1 = 1$ and $\beta_2 = 0$.
The predictor variables are independent and normally distributed with zero mean and
standard deviation one, and the simulated noise standard deviation is $\sigma_{\text{true}} = 0.5$.
The R-sens sensitivities for both variables
are shown in the bottom part of Figure~\ref{fig:lm_illustr}.
The red color
depicts the predictive distribution $p(y | x_1, x_2 = 0)$ (top) and
R-sens for $x_1$ (bottom).
The R-sens value is dominated by the contribution from the first summand in equation~(\ref{eq:2der_lm}),
where the Fisher information weighs
down the sensitivity at the edges of the data because of the larger uncertainty.
The blue color
shows the predictive distribution $p(y | x_2, x_1 = 0)$ and
R-sens for $x_2$.
Now the first summand in equation~(\ref{eq:2der_lm}) is almost zero
because $\widehat{\beta}_2$ is so small.
In this case, the second term dominates
because there is still a significant amount of epistemic uncertainty in the model.
Comparing R-sens for $x_1$ and $x_2$ illustrates the two different ways
that R-sens takes uncertainty into account.

Note that R-sens is model-agnostic in the sense that it does not take into account the fact that
the prediction function is constrained to be linear in the example above.
In addition, 
the different terms
in equation~(\ref{eq:kldiff}) may not have such clear interpretations
for other likelihoods or models.
For example, in a binary classification task, the posterior
predictive distribution can be considered a Bernoulli
distribution which has only a single parameter.
Nevertheless, the principle of taking into account
the uncertainty through the predictive distribution still holds.

\subsection{Applicability}

The requirements for using the proposed R-sens and R-sens$_2$ methods are that we must have
an analytical representation of the predictive distribution conditioned on the predictor variables, and
that the derivatives of its parameters with respect to the predictors must be available.
This somewhat restricts their applicability, but
computational tools such as automatic differentiation make practical implementation easier~\citep{baydin2018automatic}.
There are no restrictions to the support of the predictive
distribution, and
the methods are thus applicable to many learning tasks.
However, because the methods adjust sensitivity based on the model
uncertainty, the
results may be misleading if the model lacks proper uncertainty quantification.
We thus recommend using the methods for probabilistic models where
the uncertainty is properly taken into consideration.
In practice, the methods are
most useful when the number of observations is relatively small and there is a lot of uncertainty about the
parameters of the model.

Because the proposed methods measure the importance of predictor variables locally, they
are most useful for nonlinear and complex models.
For example, they can be useful for sensitivity analysis with Gaussian process models, which
can represent flexible nonlinear functions with interactions, but still
have good uncertainty quantification~\citep{ohagan1978curve,mackay1998introduction,neal1998regression,rasmussen2006gaussian}.
Moreover, the predictive distribution is available in a parametric form, although for certain likelihoods
some approximations are required.
In the supplementary material we show the derivatives
required for the R-sens and R-sens$_2$ methods
for Gaussian processes and commonly used likelihoods.

The added computational expense of the R-sens and R-sens$_2$ methods compared to just differentiating
the mean prediction depends on the used model. For many
models, the cost is not significant compared to
the cost of inference.

\subsection{Finite Difference Approximation}

\citet{paananen2019variable} propose a sensitivity analysis method abbreviated KL, which is a finite difference like method that evaluates
the Kullback-Leibler divergence of predictive distributions
when the predictor variables are perturbed.
If we set $\alpha = 1$ (where R\'{e}nyi divergence equals the Kullback-Leibler divergence), we show that
the KL method is approximately equivalent to the second order Taylor approximation of R-sens
\begin{nalign*}
& \text{R-sens} (\mathbf{x}^{*},x_d, \alpha = 1)  \approx \\
& \frac{\sqrt{2 \mathcal{D}_{1} [ p (y^{*} | \boldsymbol{\lambda}^{*} (\mathbf{x}^{*})) || p (y^{*} | \boldsymbol{\lambda}^{*} (\mathbf{x}^{**}))]}}{|x_d^{**} - x_d^{*}|} .
\end{nalign*}
Here, $\mathbf{x}^{**}$ is equivalent to $\mathbf{x}^{*}$ with predictor $x_d$ perturbed,
and $\mathcal{D}_{1}$ denotes the Kullback-Leibler divergence.
Due to space constraints, we show the full derivation in the supplementary material.
The benefit of the finite difference approximation is generality as it requires only 
an analytical representation of the predictive distribution but not
the derivatives of
its parameters with respect to the predictors.
However, using R-sens
avoids numerical errors related to finite difference and
is easier because the selection of the perturbation size is avoided.
For an appropriately chosen perturbation, the
two equations produce practically identical results up to a small numerical error.

Our proposed R-sens$_2$ measure is not directly approximable with finite differences in the same way
as R-sens.
This is because it would require second-order finite differences, but the
first-order finite difference using Kullback-Leibler divergence already reduces the predictive distribution into a single number.
\citet{riihimaki2010analysing} perturb
two predictor variables at a time with a unit length perturbation and measure
the change in predictions by Kullback-Leibler divergence.
For an infinitesimal perturbation this would be equivalent to
a directional derivative instead of the cross-derivative in the R-sens$_2$ method that is required
to properly assess interaction effects.

\section{Experiments} \label{sec:experiments}

In this section, we demonstrate the practical utility of the methods
discussed in Section~\ref{sec:method} for identifying
important predictor variables and interactions in nonlinear models.
First, we
evaluate different variable importance methods on simulated data using
a hypothetical predictive function. This way we can control the quality
of the model fit and limit the comparison strictly to the variable importance methods.
Second, we will use Gaussian process models to evaluate ranking of
main effects and interactions on both simulated and real data.

We compare the R-sens and R-sens$_2$ measures to several alternative variable importance methods:
1) Expected absolute derivative (EAD) or expected absolute Hessian (EAH), which
correspond to R-sens and R-sens$_2$ without predictive uncertainty,
2) Absolute expected derivative (AED) or absolute expected Hessian (AEH) that
take the expectation over $\mathbf{x}$ inside the absolute value~\citep{cui2019recovering},
3) Average predictive comparison (APC)~\citep{gelman2007average},
4) Shapley values~\citep{shapley1953value,vstrumbelj2014explaining,lundberg2018consistent},
5) Partial dependence based importance (PD)~\citep{greenwell2018simple},
6) Permutation feature importance (PFI)~\citep{fisher2019all},
7) Variance of the predictive mean (VAR)~\citep{paananen2019variable}, and
8) H-statistic~\citep{friedman2008predictive}.
We omit comparison to the KL method of~\citet{paananen2019variable} because it is equivalent to R-sens up to numerical accuracy.
We still show the results of their VAR
method, which has no direct connection to R-sens.
We have used the methods such that their computational cost is approximately equivalent.
In the supplementary
material, we detail the practical computational cost of the compared methods.

For assessing the global importance of predictor variables or pairs of variables, we aggregate
the local R-sens and R-sens$_2$ sensitivity measures
over the empirical distribution of the predictors.
Using the global measures, we can rank the predictor variables
or pairwise interactions.

\subsection{Simulated Individual Effects} \label{sec:simul}

In the first experiment we compare different methods for ranking individual predictors
based on their importance.
We simulate $200$ observations from $10$ predictors, and construct the target variable $y$
as a sum of $10$ effects with added Gaussian noise
\begin{nalign*}
x_i & \sim p_{x_i}(x_i), \, \,  i = 1, \dots , 10 , \\
y & = f_{\text{true}} (\mathbf{x}) = \sum_{i = 1}^{10} A_i f_{\text{true},i}(x_i) + \varepsilon .
\end{nalign*}
The shape of each effect $f_{\text{true},i}(x_i)$ is the same for all $i$,
but they have different strengths varying from $A_1 = 1$ to $A_{10} = 10$.
We consider
6 different experiments with different function shapes.
By considering only a single effect shape for each experiment, we
can unambiguously define the true importance of each predictor.
We also repeat the experiment with 4 different distributions for the predictors.

When evaluating the ranking methods, we first use the true data generating function $f_{\text{true}} (\mathbf{x})$
as the mean prediction of the model. To simulate the uncertainty of the prediction model, we
set the predictive distribution as Gaussian whose variance increases quadratically as distance from the mean
of the data increases.
Using the true data generating function allows us to strictly compare the
ranking methods without being obfuscated by a non-optimal model fit.
Because all of the compared methods use the mean prediction for ranking the predictors, using
the true data generating function does not favour any single method over the others.

To consider the effect of taking uncertainty into account in the ranking, we
also consider an imperfect version of the ground-truth model, where each
term $ A_i f_{\text{true},i}(x_i)$ is multiplied with a term $(|A_{\text{bias},i}| |x_i|^3 + 1)$ where
$A_{\text{bias},i}$ is drawn from a normal distribution with mean $0$ and standard deviation $0.02$.
This simulates a situation where the model is correct where the uncertainty is small, but
is biased at the edges of the data where the uncertainty is larger.

In Table~\ref{tab:truemaineff} we show the results of different ranking methods
for 6 different functions $f_{\text{true}} (\mathbf{x})$.
Here, the distributions $p_{x_i}$ of the predictors are 
independent Student-$t_3$ distributions.
In the supplementary material, we show the results of the experiment with 
three alternative distributions $p_{x_i}$.
The results are generated from 500 independent repetitions. In each repetition, the 10 predictors
are ranked in importance from $1$ to $10$.
For each data realisation, we compute the average error in the ranks across the predictors with respect to the true ranking,
and compare that error to the ranking error of R-sens. A negative number thus means
that the error is on average smaller than for R-sens.
Table~\ref{tab:truemaineff} reports the mean and $95\%$ uncertainty invervals
of the comparative ranking errors across the 500 independent data realisations.

\begin{table*}[tbp]
\centering
\caption{Average relative errors in rankings compared to R-sens and $95 \%$ uncertainty intervals from 500 data realisations in the simulated example of Section~\ref{sec:simul}.
A negative value indicates better ranking than R-sens.}
\scalebox{0.9}{
\begin{tabular}{ c }
 \, \\
 \textbf{Ground-truth models}  \\
\end{tabular}
}
\scalebox{0.9}{
\begin{tabular}[tbp]{ c c c c c c c c c c }
\toprule
\multicolumn{2}{c}{Function $f_{\text{true},i} (x)$} & R-sens & EAD & AED & APC & SHAP & PD & PFI & VAR  \\
\midrule
\begin{minipage}[c]{6mm}
\centering
    \includegraphics[width=6mm]{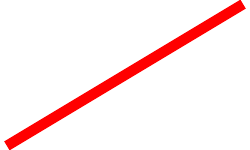}
\end{minipage} & $x$  & 
$ \mathbf{0}$ & $\mathbf{0.0 \pm 0.0}$ & $\mathbf{0.0 \pm 0.0}$ & $\mathbf{0.0 \pm 0.0}$ & $2.3 \pm 0.2$ & $1.1 \pm 0.1$ & $3.6 \pm 0.2$ & $3.9 \pm 0.2$
 \\
\hline
\begin{minipage}[c]{6mm}
\centering
    \includegraphics[width=6mm]{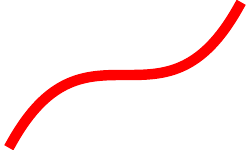}
\end{minipage} & $x^3$ & 
$ \mathbf{0}$ & $2.0 \pm 0.2$ & $2.0 \pm 0.2$ & $9.5 \pm 0.5$ & $10.0 \pm 0.4$ & $1.2 \pm 0.3$ & $15.8 \pm 0.5$ & $5.7 \pm 0.3$
 \\
\hline
\begin{minipage}[c]{6mm}
\centering
    \includegraphics[width=6mm]{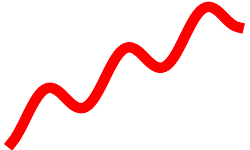}
\end{minipage} &$x + \cos (3 x)$  & 
$0$ & $ \mathbf{-0.1 \pm 0.0}$ & $4.0 \pm 0.2$ & $5.9 \pm 0.3$ & $1.7 \pm 0.2$ & $1.8 \pm 0.2$ & $2.8 \pm 0.2$ & $3.0 \pm 0.2$
\\
\hline
\begin{minipage}[c]{6mm}
\centering
    \includegraphics[width=6mm]{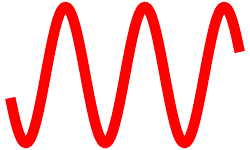}
\end{minipage} &$\sin (3 x)$  & 
$\mathbf{0}$ & $ \mathbf{0.0 \pm 0.0}$ & $20.3 \pm 0.6$ & $11.5 \pm 0.4$ & $0.4 \pm 0.1$ & $0.3 \pm 0.1$ & $0.4 \pm 0.1$ & $8.3 \pm 0.5$
\\
\hline
\begin{minipage}[c]{6mm}
\centering
    \includegraphics[width=6mm]{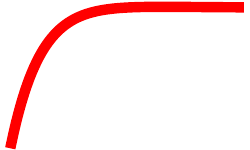}
\end{minipage} &$x \exp (-x)$  & 
$0$ & $0.6 \pm 0.1$ & $0.6 \pm 0.1$ & $0.7 \pm 0.3$ & $1.1 \pm 0.2$ & $ \mathbf{-16.5 \pm 0.7}$ & $5.6 \pm 0.7$ & $-4.6 \pm 0.7$
\\
\hline
\begin{minipage}[c]{6mm}
\centering
    \includegraphics[width=6mm]{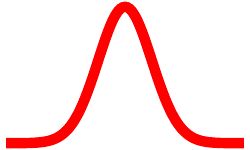}
\end{minipage} &$ \exp (-x^2)$  & 
$\mathbf{0}$ & $\mathbf{0.0 \pm 0.0}$ & $20.5 \pm 0.6$ & $9.3 \pm 0.3$ & $0.3 \pm 0.1$ & $ \mathbf{-0.1 \pm 0.1}$ & $0.3 \pm 0.1$ & $\mathbf{0.0 \pm 0.1}$
  \\
\bottomrule
\multicolumn{10}{c}{\textbf{Imperfect models}} \\
\toprule
\multicolumn{2}{c}{Function $f_{\text{true},i} (x)$} & R-sens & EAD & AED & APC & SHAP & PD & PFI & VAR  \\

\midrule
\begin{minipage}[c]{6mm}
\centering
    \includegraphics[width=6mm]{figs/boxplot1.pdf}
\end{minipage} & $x$  & 
$ \mathbf{0}$ & $2.6 \pm 0.5$ & $2.7 \pm 0.5$ & $10.8 \pm 0.7$ & $20.1 \pm 0.7$ & $22.4 \pm 0.9$ & $21.3 \pm 0.7$ & $1.1 \pm 0.4$
\\
\hline
\begin{minipage}[c]{6mm}
\centering
    \includegraphics[width=6mm]{figs/boxplot2.pdf}
\end{minipage} & $x^3$&
$0$ & $1.4 \pm 0.5$ & $1.7 \pm 0.5$ & $6.7 \pm 0.7$ & $9.7 \pm 0.8$ & $12.3 \pm 0.9$ & $9.7 \pm 0.8$ & $ \mathbf{-4.4 \pm 0.5}$ \\
\hline
\begin{minipage}[c]{6mm}
\centering
    \includegraphics[width=6mm]{figs/boxplot3.pdf}
\end{minipage} &$x + \cos (3 x)$  & 
$ \mathbf{0}$ & $2.6 \pm 0.4$ & $6.7 \pm 0.5$ & $14.0 \pm 0.6$ & $23.4 \pm 0.7$ & $25.1 \pm 0.9$ & $24.8 \pm 0.6$ & $4.0 \pm 0.4$
\\
\hline
\begin{minipage}[c]{6mm}
\centering
    \includegraphics[width=6mm]{figs/boxplot5.pdf}
\end{minipage} &$\sin (3 x)$  & 
$ \mathbf{0}$ & $2.1 \pm 0.3$ & $18.9 \pm 0.6$ & $10.5 \pm 0.5$ & $14.8 \pm 0.6$ & $5.3 \pm 0.9$ & $20.8 \pm 0.7$ & $1.0 \pm 0.3$
\\
\hline
\begin{minipage}[c]{6mm}
\centering
    \includegraphics[width=6mm]{figs/boxplot6.pdf}
\end{minipage} &$x \exp (-x)$  & 
$0$ & $0.2 \pm 0.3$ & $0.5 \pm 0.3$ & $4.3 \pm 0.8$ & $3.9 \pm 0.8$ & $5.2 \pm 1.0$ & $4.0 \pm 0.8$ & $ \mathbf{-2.7 \pm 0.8}$
\\
\hline
\begin{minipage}[c]{6mm}
\centering
    \includegraphics[width=6mm]{figs/boxplot7.pdf}
\end{minipage} &$ \exp (-x^2)$  & 
$\mathbf{0}$ & $ \mathbf{0.0 \pm 0.0}$ & $20.6 \pm 0.6$ & $9.1 \pm 0.3$ & $0.3 \pm 0.1$ & $\mathbf{0.0 \pm 0.1}$ & $0.3 \pm 0.1$ & $\mathbf{0.0 \pm 0.1}$
\\
\bottomrule

\end{tabular}
}
\label{tab:truemaineff}
\end{table*}

The top section of Table~\ref{tab:truemaineff} shows the ranking errors when using the ground-truth predictions function.
R-sens
and EAD are almost equivalent in many cases, but R-sens is significantly better for functions that
have a large derivative in the tails of the data ($x^3$ and $x \exp(-x)$).
Both R-sens and EAD outperform the other methods in most cases.
AED does well for function that are monotonic, but it fails badly
for non-monotonic functions. This is expected because the derivative
of these functions varies from positive to negative.

In the bottom section of Table~\ref{tab:truemaineff}
when the model's predictions are imperfect, the difference in the ranking errors
of R-sens and EAD is significantly larger in favour of R-sens.
R-sens is also consistently better than the alternative methods with just a few exceptions.
This shows that the uncertainty-aware sensitivity can be more reliable
when there is a significant amount of uncertainty.
In the supplementary material, we repeat the experiment with 
three alternative distributions $p_{x_i}$, including
independent and correlated normal distributions.
These results have similar conclusions: R-sens is mostly similar to EAD, but better
in specific situations.

\subsection{Simulated Individual and Pairwise Effects}

In the second experiment, we study how accurately different methods evaluate interactions
when the model has both main effects and interaction effects. We simulate 12
predictors and 8
main effects with different shapes and strengths, and three equally important pairwise interaction effects
which are simply the product of the two predictors, i.e. $x_d x_e$.
These are chosen such that the predictors of the first interaction do not have main effects,
one of the predictors in the second interaction has a main effect, and
both predictors of the third interaction have a main effect.
To study how many observations the different methods require
to reliably detect the true interactions in the data,
we generate data with different numbers of observations ranging from $50$ to $300$.

\begin{figure*}[htb]
\centering
\includegraphics[width=\textwidth]{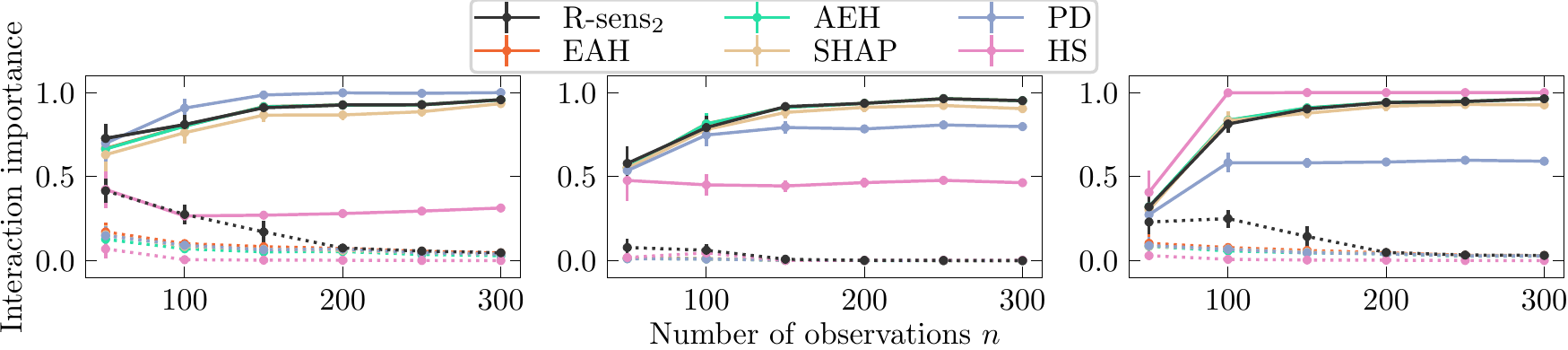}
\vspace{-5mm}
\caption{The interaction importance values given
to six variable pairs
in data sets with different numbers of observations.
The solid lines represent pairs with a true interaction, and dotted lines are pairs without
an interaction.
The left plot depicts two pairs where
both variables in the pairs have a main effect.
In the middle plot, only one variable in the pairs has a main effect, and
in the right plot neither variable has a main effect.
In all plots, the ideal values would be 1 and 0 for the solid and dotted lines, respectively. The error bars represent $95 \%$ uncertainty
intervals for the means from 50 simulated data sets.} \label{fig:toy_n}
\end{figure*}

In Figure~\ref{fig:toy_n}, we plot the importance values averaged from 50 simulations for the three interacting variable pairs
as well as three variable pairs without an interaction effect.
The solid lines represent pairs with a true interaction, and the dotted lines are pairs without
an interaction. In the left plot, both variables in the pairs have a main effect.
In the middle plot, only one variable in the pairs has a main effect, and
in the right plot neither variable has a main effect.
For each of the 50 simulations, the interaction importance values are scaled so that the maximum given by each method is one.
Thus, the ideal value is 1 for the solid lines and 0 for the dotted lines.

Figure~\ref{fig:toy_n} shows that even when increasing the number of observations,
the HS method over-emphasizes the variable pair where neither variable
has a main effect (right plot), whereas the PD method over-emphasizes
the variable pair where both variables
have a main effect (left plot).
The other methods correctly identify the interactions as equally relevant when increasing the number of observations.
For the true interactions (solid lines), EAH and R-sens$_2$ are almost indistinguishable, but
R-sens$_2$ gives higher importance to the nonexistent interactions (dotted lines) when
there is significant uncertainty because the number of observations is small.

\subsection{Benchmark Data Sets}

In real data experiments, we focus on assessing the performance of the pairwise interaction method R-sens$_2$,
because the experiments of~\citet{paananen2019variable} already demonstrate the effectiveness of (the finite difference approximation of) R-sens empirically.
We use two publicly available data sets. The first is the Concrete Slump data set where
the compressive strength of concrete is predicted based on the amount of different components included~\citep{yeh2007modeling}.
The second is a Bike sharing data set, where the target variable is the hourly number of bike uses from a bicycle rental system~\citep{fanaee2014event}.
The Concrete data has 103 observations and 7 predictors.
From the Bike sharing data, we picked observations from February across two years, resulting in 1339 observations
and 6 predictors.
We model the problems using Gaussian process models with
an exponentiated quadratic covariance function and
either Gaussian (Concrete) or Poisson (Bike) likelihood.
With the Poisson likelihood, we use the Laplace approximation for the latent values,
and thus the resulting predictive distribution is an approximation but has an analytical solution.
The details of the models and the derivatives
needed for R-sens$_2$ are presented in the supplementary material.

To evaluate the plausibility of the interactions identified by the
different methods, we compare
the out-of-sample predictive performance of
models with explicit interaction terms chosen based on interactions identified by each method.
We compare the performance of the models
using cross-validation with 50 random splits into training and test sets, and log predictive density as the utility function.
The number of training observations used is $80$ in the Concrete data and $500$ in the Bike sharing data.
For each training set, the Gaussian process model with full interactions is fitted, and
the pairwise interactions are identified with R-sens$_2$ and 5 other methods.
Based on these, models with only 0 to 5 pairwise interaction terms are fitted again, and their
predictive performance is evaluated on the test data.

\begin{figure}[htb]
 \centering
\includegraphics[width=0.49\textwidth]{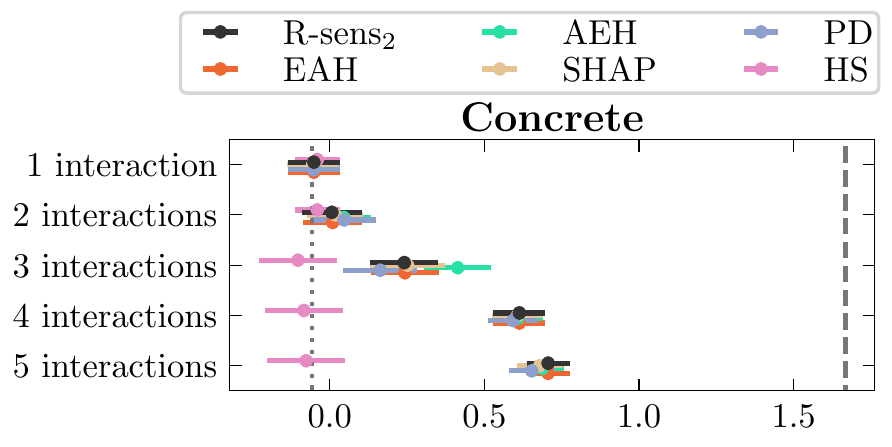}
\includegraphics[width=0.49\textwidth]{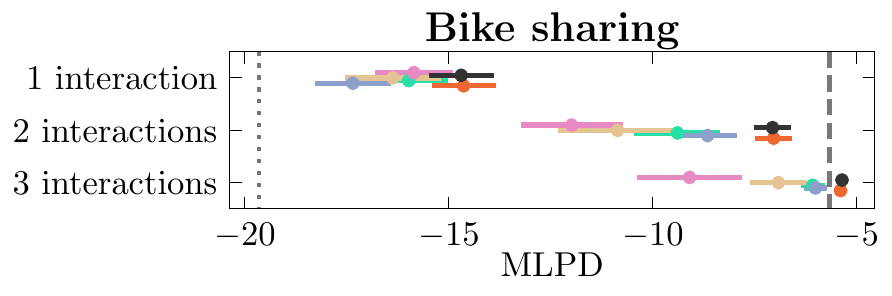}
\caption{
Mean log predictive densities (MLPDs) on independent test sets from the Concrete and Bike sharing data sets for Gaussian process models with different numbers of interactions. With each method, the interactions were identified from each training data set using the model with all interactions.
The error bars represent $95 \%$ uncertainty
intervals for the means from 50 different train-test splits.
The dotted and dashed lines represent the MLPD for models with no interactions and all interactions,
respectively.
} \label{fig:bikeshare_concrete_MLPD}
\end{figure}

The mean log predictive densities (MLPDs) across different test splits
as well as $95\%$ uncertainty intervals of the means are shown in Figure~\ref{fig:bikeshare_concrete_MLPD}.
The figure shows that in the Bike sharing data, modelling only the three strongest pairwise interactions increases the out-of-sample predictive performance to the level of the model with all possible interactions.
R-sens$_2$ does equally well compared to EAH, which both
identify more important interactions than the competing methods.
In the Concrete data set, adding even 5 pairwise interactions does not
reach the performance of the model with all interactions.
In this data there are no significant differences between the methods,
except for HS which is clearly worse than the rest.

We also evaluate the stability of the interaction rankings by computing the variability in the rankings
across 100 Bootstrap samples of the data. Table~\ref{tab:bikedata_stab} shows the entropy in the rankings of each
method across the Bootstrap samples. In both data sets, R-sens$_2$ and EAH have smaller entropies than the competing methods,
meaning that their rankings are more stable.

\begin{table}[htb]
\centering
\caption{Variability in rankings across different Bootstrap samples of the benchmark data sets.}
\scalebox{0.88}{
\begin{tabular}[tbp]{l|cccccc}
Data & R-sens$_2$ & EAH & AEH & PD & HS & SHAP \\
\midrule
Concrete & $1.86 $ &$\mathbf{1.80} $ &$1.99$ &$2.04 $ &$2.26 $ &$1.90 $  \\
Bike & $\mathbf{1.61} $ &$1.63 $ &$2.28$ &$2.16 $ &$2.61 $ &$2.58 $  \\
\end{tabular}
}
\label{tab:bikedata_stab}
\end{table}

\section{Conclusion} \label{sec:conclusions}

In this work, we presented an uncertainty-aware sensitivity analysis method
that is based
on differentiating R\'{e}nyi divergences of predictive distributions.
We showed that the method takes model uncertainty into account in a principled way
and generalises to different likelihoods.
For likelihoods in the location-scale family, the method is a direct extension
to the absolute derivative
or absolute Hessian of the mean prediction which are non-Bayesian sensitivity measures.
Even though the method generalises to different predictive distributions, we recommend using
it for models that have well calibrated uncertainty.
The proposed method requires an analytical representation of the predictive distribution
of the model, which is not available for all models.
This could be generalised further, which is a possible direction for future research.

We demonstrated empirically that the method can reliably
identify main effects as well as interactions in nonlinear models for complex data sets.
In a controlled simulation setting, we showed that using uncertainty-aware sensitivity is beneficial
in the presence of uncertainty when the used model may be wrong.
Moreover, the proposed methods were
equally good or better than previous derivative based sensitivity analysis methods in all
of the tested cases.
We can thus recommend using uncertainty-aware sensitivity analysis in modelling situations with
little data and/or lots of uncertainty.
We also demonstrated with two real data sets that our proposed method
identifies pairwise interactions in nonlinear models that, when added to a model,
improve its predictive performance. In addition, the ranking of the interactions
between different Bootstrap samples of the data has less variation compared to many alternative variable importance methods.

\subsection*{Acknowledgements} 
We thank Alejandro Catalina and Kunal Ghosh for helpful comments to improve the manuscript. We thank Mostafa Abdelrahman for GPyTorch and autodiff implementations.
We also acknowledge the computational resources provided by the Aalto Science-IT project and support by the Academy of Finland Flagship
programme: Finnish Center for Artificial Intelligence, FCAI.

\bibliography{rsens}

\begin{thebibliography}{}

\bibitem[Baydin et~al., 2018]{baydin2018automatic}
Baydin, A.~G., Pearlmutter, B.~A., Radul, A.~A., and Siskind, J.~M. (2018).
\newblock Automatic differentiation in machine learning: a survey.
\newblock {\em Journal of machine learning research}, 18(153).

\bibitem[Cacuci, 2003]{cacuci2003sensitivity}
Cacuci, D.~G. (2003).
\newblock {\em Sensitivity and Uncertainty Analysis, volume I: Theory}.
\newblock Boca Raton: Chapman and Hall/CRC.

\bibitem[Cacuci et~al., 2005]{cacuci2005sensitivity}
Cacuci, D.~G., Ionescu-Bujor, M., and Navon, I.~M. (2005).
\newblock {\em Sensitivity and uncertainty analysis, volume II: applications to
  large-scale systems}.
\newblock CRC press.

\bibitem[Cui et~al., 2020]{cui2019recovering}
Cui, T., Marttinen, P., and Kaski, S. (2020).
\newblock Learning global pairwise interactions with bayesian neural networks.
\newblock {\em Proceedings of the 24th European Conference on Artificial
  Intelligence (ECAI 2020)}.

\bibitem[Czernichow, 1996]{czernichow1996architecture}
Czernichow, T. (1996).
\newblock Architecture selection through statistical sensitivity analysis.
\newblock In {\em International Conference on Artificial Neural Networks},
  pages 179--184. Springer.

\bibitem[Dorizzi, 1996]{dorizzi1996variable}
Dorizzi, B. (1996).
\newblock Variable selection using generalized {RBF} networks: Application to
  the forecast of the {French} {T-bonds}.
\newblock {\em Proceedings of IEEE-IMACS'96, Lille, France}.

\bibitem[Fanaee-T and Gama, 2014]{fanaee2014event}
Fanaee-T, H. and Gama, J. (2014).
\newblock Event labeling combining ensemble detectors and background knowledge.
\newblock {\em Progress in Artificial Intelligence}, 2(2-3):113--127.

\bibitem[Fisher et~al., 2019]{fisher2019all}
Fisher, A., Rudin, C., and Dominici, F. (2019).
\newblock All models are wrong, but many are useful: Learning a variable's
  importance by studying an entire class of prediction models simultaneously.
\newblock {\em Journal of Machine Learning Research}, 20(177):1--81.

\bibitem[Friedman, 2001]{friedman2001greedy}
Friedman, J.~H. (2001).
\newblock Greedy function approximation: a gradient boosting machine.
\newblock {\em Annals of statistics}, pages 1189--1232.

\bibitem[Friedman et~al., 2008]{friedman2008predictive}
Friedman, J.~H., Popescu, B.~E., et~al. (2008).
\newblock Predictive learning via rule ensembles.
\newblock {\em The Annals of Applied Statistics}, 2(3):916--954.

\bibitem[Fujimoto et~al., 2006]{fujimoto2006axiomatic}
Fujimoto, K., Kojadinovic, I., and Marichal, J.-L. (2006).
\newblock Axiomatic characterizations of probabilistic and
  cardinal-probabilistic interaction indices.
\newblock {\em Games and Economic Behavior}, 55(1):72--99.

\bibitem[Gelman and Pardoe, 2007]{gelman2007average}
Gelman, A. and Pardoe, I. (2007).
\newblock Average predictive comparisons for models with nonlinearity,
  interactions, and variance components.
\newblock {\em Sociological Methodology}, 37(1):23--51.

\bibitem[Goldstein et~al., 2015]{goldstein2015peeking}
Goldstein, A., Kapelner, A., Bleich, J., and Pitkin, E. (2015).
\newblock Peeking inside the black box: Visualizing statistical learning with
  plots of individual conditional expectation.
\newblock {\em Journal of Computational and Graphical Statistics},
  24(1):44--65.

\bibitem[Greenwell et~al., 2018]{greenwell2018simple}
Greenwell, B.~M., Boehmke, B.~C., and McCarthy, A.~J. (2018).
\newblock A simple and effective model-based variable importance measure.
\newblock {\em arXiv preprint arXiv:1805.04755}.

\bibitem[Guidotti et~al., 2018]{guidotti2018survey}
Guidotti, R., Monreale, A., Ruggieri, S., Turini, F., Giannotti, F., and
  Pedreschi, D. (2018).
\newblock A survey of methods for explaining black box models.
\newblock {\em ACM computing surveys (CSUR)}, 51(5):1--42.

\bibitem[Guyon and Elisseeff, 2003]{guyon2003introduction}
Guyon, I. and Elisseeff, A. (2003).
\newblock An introduction to variable and feature selection.
\newblock {\em Journal of machine learning research}, 3(Mar):1157--1182.

\bibitem[Haussler et~al., 1997]{haussler1997mutual}
Haussler, D., Opper, M., et~al. (1997).
\newblock Mutual information, metric entropy and cumulative relative entropy
  risk.
\newblock {\em The Annals of Statistics}, 25(6):2451--2492.

\bibitem[Homma and Saltelli, 1996]{homma1996importance}
Homma, T. and Saltelli, A. (1996).
\newblock Importance measures in global sensitivity analysis of nonlinear
  models.
\newblock {\em Reliability Engineering \& System Safety}, 52(1):1--17.

\bibitem[IM, 1993]{im1993sensitivity}
IM, S. (1993).
\newblock Sensitivity estimates for nonlinear mathematical models.
\newblock {\em Math. Model. Comput. Exp}, 1(4):407--414.

\bibitem[Kendall and Gal, 2017]{kendall2017uncertainties}
Kendall, A. and Gal, Y. (2017).
\newblock What uncertainties do we need in {Bayesian} deep learning for
  computer vision?
\newblock In {\em Proceedings of the 31st International Conference on Neural
  Information Processing Systems}, pages 5580--5590.

\bibitem[Kullback, 1959]{kullback1959statistics}
Kullback, S. (1959).
\newblock Statistics and information theory.
\newblock {\em J. Wiley and Sons, New York}.

\bibitem[Kullback and Leibler, 1951]{kullback1951information}
Kullback, S. and Leibler, R.~A. (1951).
\newblock On information and sufficiency.
\newblock {\em The annals of mathematical statistics}, 22(1):79--86.

\bibitem[Leray and Gallinari, 1999]{leray1999feature}
Leray, P. and Gallinari, P. (1999).
\newblock Feature selection with neural networks.
\newblock {\em Behaviormetrika}, 26(1):145--166.

\bibitem[Lundberg et~al., 2018]{lundberg2018consistent}
Lundberg, S.~M., Erion, G.~G., and Lee, S.-I. (2018).
\newblock Consistent individualized feature attribution for tree ensembles.
\newblock {\em arXiv preprint arXiv:1802.03888}.

\bibitem[MacKay, 1998]{mackay1998introduction}
MacKay, D.~J. (1998).
\newblock Introduction to {Gaussian} processes.
\newblock In Bishop, J., editor, {\em Neural Networks and Machine Learning},
  pages 133--166. Springer Verlag.

\bibitem[Minka, 2001]{minka2001family}
Minka, T.~P. (2001).
\newblock {\em A family of algorithms for approximate Bayesian inference}.
\newblock PhD thesis, Massachusetts Institute of Technology.

\bibitem[Neal, 1998]{neal1998regression}
Neal, R. (1998).
\newblock Regression and classification using {Gaussian} process priors (with
  discussion).
\newblock In Bernardo, J., Berger, J., Dawid, A., and Smith, A., editors, {\em
  Bayesian statistics}, volume~6, pages 475--501. Oxford University Press.

\bibitem[Oakley and O'Hagan, 2004]{oakley2004probabilistic}
Oakley, J.~E. and O'Hagan, A. (2004).
\newblock Probabilistic sensitivity analysis of complex models: a {Bayesian}
  approach.
\newblock {\em Journal of the Royal Statistical Society: Series B (Statistical
  Methodology)}, 66(3):751--769.

\bibitem[O'Hagan, 1978]{ohagan1978curve}
O'Hagan, A. (1978).
\newblock Curve fitting and optimal design for prediction.
\newblock {\em Journal of the Royal Statistical Society: Series B
  (Methodological)}, 40(1):1--24.

\bibitem[O'Hagan, 2004]{o2004dicing}
O'Hagan, T. (2004).
\newblock Dicing with the unknown.
\newblock {\em Significance}, 1(3):132--133.

\bibitem[Opper and Winther, 2000]{Opper+Winther:2000}
Opper, M. and Winther, O. (2000).
\newblock {G}aussian processes for classification: Mean-field algorithms.
\newblock {\em Neural Computation}, 12(11):2655--2684.

\bibitem[Paananen et~al., 2019]{paananen2019variable}
Paananen, T., Piironen, J., Andersen, M.~R., and Vehtari, A. (2019).
\newblock Variable selection for {Gaussian} processes via sensitivity analysis
  of the posterior predictive distribution.
\newblock volume~89 of {\em Proceedings of Machine Learning Research}, pages
  1743--1752.

\bibitem[Rasmussen, 2003]{rasmussen2003gaussian}
Rasmussen, C. (2003).
\newblock Gaussian processes to speed up hybrid monte carlo for expensive
  bayesian integrals.
\newblock {\em Bayesian statistics}, 7:651--659.

\bibitem[Rasmussen and Williams, 2006]{rasmussen2006gaussian}
Rasmussen, C.~E. and Williams, C.~K. (2006).
\newblock {\em Gaussian processes for machine learning}, volume~1.
\newblock MIT press Cambridge.

\bibitem[Refenes and Zapranis, 1999]{refenes1999neural}
Refenes, A.-P. and Zapranis, A. (1999).
\newblock Neural model identification, variable selection and model adequacy.
\newblock {\em Journal of Forecasting}, 18(5):299--332.

\bibitem[R{\'e}nyi et~al., 1961]{renyi1961measures}
R{\'e}nyi, A. et~al. (1961).
\newblock On measures of entropy and information.
\newblock In {\em Proceedings of the Fourth Berkeley Symposium on Mathematical
  Statistics and Probability, Volume 1: Contributions to the Theory of
  Statistics}. The Regents of the University of California.

\bibitem[Riihim{\"a}ki et~al., 2010]{riihimaki2010analysing}
Riihim{\"a}ki, J., Sund, R., and Vehtari, A. (2010).
\newblock Analysing the length of care episode after hip fracture: a
  nonparametric and a parametric {Bayesian} approach.
\newblock {\em Health care management science}, 13(2):170--181.

\bibitem[Ruck et~al., 1990]{ruck1990feature}
Ruck, D.~W., Rogers, S.~K., and Kabrisky, M. (1990).
\newblock Feature selection using a multilayer perceptron.
\newblock {\em Journal of Neural Network Computing}, 2(2):40--48.

\bibitem[Saltelli, 2002]{saltelli2002sensitivity}
Saltelli, A. (2002).
\newblock Sensitivity analysis for importance assessment.
\newblock {\em Risk analysis}, 22(3):579--590.

\bibitem[Shao, 2006]{shao2006mathematical}
Shao, J. (2006).
\newblock {\em Mathematical statistics: exercises and solutions}.
\newblock Springer Science \& Business Media.

\bibitem[Shapley, 1953]{shapley1953value}
Shapley, L.~S. (1953).
\newblock A value for n-person games.
\newblock {\em Contributions to the Theory of Games}, 2(28):307--317.

\bibitem[Simonyan et~al., 2013]{simonyan2013deep}
Simonyan, K., Vedaldi, A., and Zisserman, A. (2013).
\newblock Deep inside convolutional networks: Visualising image classification
  models and saliency maps.
\newblock {\em arXiv preprint arXiv:1312.6034}.

\bibitem[Solak et~al., 2003]{solak2003derivative}
Solak, E., Murray-Smith, R., Leithead, W.~E., Leith, D.~J., and Rasmussen,
  C.~E. (2003).
\newblock Derivative observations in {Gaussian} process models of dynamic
  systems.
\newblock In {\em Advances in neural information processing systems}, pages
  1057--1064.

\bibitem[{\v{S}}trumbelj and Kononenko, 2014]{vstrumbelj2014explaining}
{\v{S}}trumbelj, E. and Kononenko, I. (2014).
\newblock Explaining prediction models and individual predictions with feature
  contributions.
\newblock {\em Knowledge and information systems}, 41(3):647--665.

\bibitem[Sundararajan et~al., 2017]{sundararajan2017axiomatic}
Sundararajan, M., Taly, A., and Yan, Q. (2017).
\newblock Axiomatic attribution for deep networks.
\newblock In {\em Proceedings of the 34th International Conference on Machine
  Learning}, pages 3319--3328.

\bibitem[Van~Erven and Harremos, 2014]{van2014renyi}
Van~Erven, T. and Harremos, P. (2014).
\newblock R{\'e}nyi divergence and {Kullback}-{Leibler} divergence.
\newblock {\em IEEE Transactions on Information Theory}, 60(7):3797--3820.

\bibitem[Walker, 1969]{walker1969asymptotic}
Walker, A.~M. (1969).
\newblock On the asymptotic behaviour of posterior distributions.
\newblock {\em Journal of the Royal Statistical Society: Series B
  (Methodological)}, 31(1):80--88.

\bibitem[Williams and Barber, 1998]{williams1998bayesian}
Williams, C.~K. and Barber, D. (1998).
\newblock Bayesian classification with {Gaussian} processes.
\newblock {\em IEEE Transactions on Pattern Analysis and Machine Intelligence},
  20(12):1342--1351.

\bibitem[Yeh, 2007]{yeh2007modeling}
Yeh, I.-C. (2007).
\newblock Modeling slump flow of concrete using second-order regressions and
  artificial neural networks.
\newblock {\em Cement and concrete composites}, 29(6):474--480.

\bibitem[Zeiler and Fergus, 2014]{zeiler2014visualizing}
Zeiler, M.~D. and Fergus, R. (2014).
\newblock Visualizing and understanding convolutional networks.
\newblock In {\em European conference on computer vision}, pages 818--833.
  Springer.

\end{thebibliography}

\onecolumn

\section{R-SENS AND R-SENS$_2$ DERIVATIONS}

\subsection{R-sens}

\begin{nalign}
&  \frac{\partial^2 \mathcal{D}_{\alpha} [ \, p  (y^{*} | \boldsymbol{\lambda}^{*}(\mathbf{x}^{*}) ) || p  (y^{*} | \boldsymbol{\lambda}^{*}(\mathbf{x}^{**} ) )]}{(\partial x_d^{**})^2} \bigg|_{ \mathbf{x}^{**} =  \mathbf{x}^{*}} \\
= & \left (\frac{\partial^2 \boldsymbol{\lambda}^{*} (\mathbf{x}^{*})}{\partial (x_d^{*})^2} \right )^{T}   
\left (\frac{\mathcal{D}_{\alpha} [ \, p  (y^{*} | \boldsymbol{\lambda}^{*} (\mathbf{x}^{*}) ) || p  (y^{*} | \boldsymbol{\lambda}^{*} (\mathbf{x}^{**})) ]}{\partial \boldsymbol{\lambda}^{*} (\mathbf{x}^{**})} \right )\bigg|_{ \mathbf{x}^{**} =  \mathbf{x}^{*}}
+ \\
& \left (\frac{\partial \boldsymbol{\lambda}^{*} (\mathbf{x}^{*})}{\partial x_d^{*}} \right )^{T}   
\mathbf{H}_{\boldsymbol{\lambda}^{*} (\mathbf{x}^{**})} ( \mathcal{D}_{\alpha} [ \, p  (y^{*} | \boldsymbol{\lambda}^{*} (\mathbf{x}^{*}) ) || p  (y^{*} | \boldsymbol{\lambda}^{*} (\mathbf{x}^{**})) ] )
\left (\frac{\partial \boldsymbol{\lambda}^{*} (\mathbf{x}^{**})}{\partial x_d^{**}} \right ) \bigg|_{ \mathbf{x}^{**} =  \mathbf{x}^{*}} , \\
= & \, \mathbf{0} + \left (\frac{\partial \boldsymbol{\lambda}^{*} (\mathbf{x}^{*})}{\partial x_d^{*}} \right )^{T}   
\mathbf{H}_{\boldsymbol{\lambda}^{*} (\mathbf{x}^{**})} ( \mathcal{D}_{\alpha} [ \, p  (y^{*} | \boldsymbol{\lambda}^{*} (\mathbf{x}^{*}) ) || p  (y^{*} | \boldsymbol{\lambda}^{*} (\mathbf{x}^{**})) ] )
\left (\frac{\partial \boldsymbol{\lambda}^{*} (\mathbf{x}^{**})}{\partial x_d^{**}} \right ) \bigg|_{ \mathbf{x}^{**} =  \mathbf{x}^{*}} .
\end{nalign}

\subsection{R-sens$_2$}

Here, we make the approximation that third and fourth derivatives of the R\'{e}nyi divergence are
zero.
Let us start from the previous identity

\begin{nalign}
& \frac{\partial^2 \mathcal{D}_{\alpha} [ \, p  (y^{*} | \boldsymbol{\lambda}^{*}(\mathbf{x}^{*}) ) || p  (y^{*} | \boldsymbol{\lambda}^{*}(\mathbf{x}^{**} ) )]}{(\partial x_d^{**})^2} \bigg|_{ \mathbf{x}^{**} =  \mathbf{x}^{*}} \\
= & \left (\frac{\partial \boldsymbol{\lambda}^{*} (\mathbf{x}^{*})}{\partial x_d^{*}} \right )^{T}   
\mathbf{H}_{\boldsymbol{\lambda}^{*} (\mathbf{x}^{**})} ( \mathcal{D}_{\alpha} [ \, p  (y^{*} | \boldsymbol{\lambda}^{*} (\mathbf{x}^{*}) ) || p  (y^{*} | \boldsymbol{\lambda}^{*} (\mathbf{x}^{**})) ] )
\left (\frac{\partial \boldsymbol{\lambda}^{*} (\mathbf{x}^{**})}{\partial x_d^{**}} \right ) \bigg|_{ \mathbf{x}^{**} =  \mathbf{x}^{*}}\\
= & \, \sum_{k = 1}^M \sum_{l = 1}^M  \frac{\partial^2 \mathcal{D}_{\alpha} [ \, p  (y^{*} | \boldsymbol{\lambda}^{*} ) || p  (y^{*} | \boldsymbol{\lambda}^{*} )] }{\partial \lambda_k^{*} \partial \lambda_l^{*}}
\frac{\partial \lambda_k^{*}}{\partial x_d^{*}}
\frac{\partial \lambda_l^{*}}{\partial x_d^{*}} \bigg|_{ \mathbf{x}^{**} =  \mathbf{x}^{*}}
.
\end{nalign}
Then differentiating with respect to $x_e$ gives the equality
\begin{nalign}
& \frac{\partial^3 \mathcal{D}_{\alpha} [ \, p  (y^{*} | \boldsymbol{\lambda}^{*}(\mathbf{x}^{*}) ) || p  (y^{*} | \boldsymbol{\lambda}^{*}(\mathbf{x}^{**} ) ]}{(\partial x_d^{**})^2 \partial x_e^{**}} \bigg|_{ \mathbf{x}^{**} =  \mathbf{x}^{*}} 
=  \sum_{k = 1}^M \sum_{l = 1}^M  \sum_{m = 1}^M  \frac{\partial^3 \mathcal{D}_{\alpha} [ \, p  (y^{*} | \boldsymbol{\lambda}^{*} ) || p  (y^{*} | \boldsymbol{\lambda}^{*} ) ] }{\partial \lambda_k^{*} \partial \lambda_l^{*} \partial \lambda_m^{*}}
\frac{\partial \lambda_k^{*}}{\partial x_d^{*}}
\frac{\partial \lambda_l^{*}}{\partial x_d^{*}}
\frac{\partial \lambda_m^{*}}{\partial x_d^{*}} \\
& +
\sum_{k = 1}^M \sum_{l = 1}^M  \frac{\partial^2 \mathcal{D}_{\alpha} [ \, p  (y^{*} | \boldsymbol{\lambda}^{*} ) || p  (y^{*} | \boldsymbol{\lambda}^{*} ] ) }{\partial \lambda_k^{*} \partial \lambda_l^{*}} \left (
\frac{\partial^2 \lambda_k^{*}}{\partial x_d^{*} \partial x_e^{*}}
\frac{\partial \lambda_l^{*}}{\partial x_d^{*}}
+
\frac{\partial \lambda_k^{*}}{\partial x_d^{*}}
\frac{\partial^2 \lambda_l^{*}}{\partial x_d^{*} \partial x_e^{*}}
\right )\\
& = 
 \sum_{k = 1}^M \sum_{l = 1}^M  \frac{\partial^2 \mathcal{D}_{\alpha} [ \, p  (y^{*} | \boldsymbol{\lambda}^{*} ) || p  (y^{*} | \boldsymbol{\lambda}^{*} ) ] }{\partial \lambda_k^{*} \partial \lambda_l^{*}} \left (
\frac{\partial^2 \lambda_k^{*}}{\partial x_d^{*} \partial x_e^{*}}
\frac{\partial \lambda_l^{*}}{\partial x_d^{*}}
+
\frac{\partial \lambda_k^{*}}{\partial x_d^{*}}
\frac{\partial^2 \lambda_l^{*}}{\partial x_d^{*} \partial x_e^{*}}
\right ).
\end{nalign}
Differentiating a second time gives
\begin{nalign}
& \frac{\partial^4 \mathcal{D}_{\alpha} [ \, p  (y^{*} | \boldsymbol{\lambda}^{*}(\mathbf{x}^{*}) ) || p  (y^{*} | \boldsymbol{\lambda}^{*}(\mathbf{x}^{**} ) ) ]}{(\partial x_d^{**})^2 (\partial x_e^{**})^2} \bigg|_{ \mathbf{x}^{**} =  \mathbf{x}^{*}} \\
& = 
 \sum_{k = 1}^M \sum_{l = 1}^M  \frac{\partial^2 \mathcal{D}_{\alpha} [ \, p  (y^{*} | \boldsymbol{\lambda}^{*} ) || p  (y^{*} | \boldsymbol{\lambda}^{*} )] }{\partial \lambda_k^{*} \partial \lambda_l^{*}} \left (
\frac{\partial^3 \lambda_k^{*}}{\partial x_d^{*} \partial x_e^{*}}
\frac{\partial \lambda_l^{*}}{\partial x_d^{*}}
+
\frac{\partial^2 \lambda_k^{*}}{\partial x_d^{*} \partial x_e^{*}}
\frac{\partial^2 \lambda_l^{*}}{\partial x_d^{*}  \partial x_e^{*}}
+
\frac{\partial^2 \lambda_k^{*}}{\partial x_d^{*} \partial x_e^{*}}
\frac{\partial^2 \lambda_l^{*}}{\partial x_d^{*} \partial x_e^{*}}
+
\frac{\partial^2 \lambda_k^{*}}{\partial x_d^{*}  \partial x_e^{*}}
\frac{\partial^3 \lambda_l^{*}}{\partial x_d^{*} (\partial x_e^{*})^2}
\right ).
\end{nalign}
Dropping the third derivative terms and the factor $2$ results in 
\begin{nalign}
 \sum_{k = 1}^M \sum_{l = 1}^M  \frac{\partial^2 \mathcal{D}_{\alpha} [ \, p  (y^{*} | \boldsymbol{\lambda}^{*} ) || p  (y^{*} | \boldsymbol{\lambda}^{*} )] }{\partial \lambda_k^{*} \partial \lambda_l^{*}}
\frac{\partial^2 \lambda_k^{*}}{\partial x_d^{*} \partial x_e^{*}}
\frac{\partial^2 \lambda_l^{*}}{\partial x_d^{*}  \partial x_e^{*}}
=
\left (\frac{\partial^2 \boldsymbol{\lambda}^{*}}{\partial x_d^{*} \partial x_e^{*}} \right )^{T}   
\mathbf{H} ( \mathcal{D}_{\alpha} [ \, p  (y^{*} | \boldsymbol{\lambda}^{*} ) || p  (y^{*} | \boldsymbol{\lambda}^{*} ) ] )
\left (\frac{\partial^2 \boldsymbol{\lambda}^{*}}{\partial x_d^{*} \partial x_e^{*}} \right )
.
\end{nalign}

\clearpage

\section{FINITE DIFFERENCE APPROXIMATION OF THE KULLBACK-LEIBLER DIVERGENCE}

Consider two probability distributions,
$p (\cdot | \boldsymbol{\lambda}^{*})$ and $p (\cdot | \boldsymbol{\lambda}^{**})$
parameterised by vectors $\boldsymbol{\lambda}^{*}$ and $\boldsymbol{\lambda}^{**}$,
respectively. Keeping $\boldsymbol{\lambda}^{*}$ constant, let us make
a second-order approximation of the Kullback-Leibler divergence between the distributions
in the neighbourhood around $\boldsymbol{\lambda}^{**} = \boldsymbol{\lambda}^{*}$.
\begin{nalign*}
& \mathcal{D}_{\text{KL}} (p (\cdot | \boldsymbol{\lambda}^{*}) || p (\cdot | \boldsymbol{\lambda}^{**})) \\
& = \mathcal{D}_{\text{KL}} (p (\cdot | \boldsymbol{\lambda}^{*}) || p (\cdot | \boldsymbol{\lambda}^{**})) \bigg|_{ \boldsymbol{\lambda}^{**} =  \boldsymbol{\lambda}^{*} } 
 + \sum_{k = 1}^{M} ( \lambda_k^{**} - \lambda_k^{*}) \frac{ \partial \mathcal{D}_{\text{KL}} (p (\cdot | \boldsymbol{\lambda}^{*}) || p (\cdot | \boldsymbol{\lambda}^{**}))}{\partial \lambda_k^{**}} \bigg|_{ \boldsymbol{\lambda}^{**} =  \boldsymbol{\lambda}^{*} } \\
 & + \frac{1}{2} \sum_{k = 1}^{M} \sum_{l = 1}^{M} \Bigg [ ( \lambda_k^{**} - \lambda_k^{*}) ( \lambda_l^{**} - \lambda_l^{*}) 
  \frac{\partial^2 \mathcal{D}_{\text{KL}} (p  (\cdot | \boldsymbol{\lambda}^{*}) || p  (\cdot |  \boldsymbol{\lambda}^{**}) )}{\partial \lambda_k^{**} \partial \lambda_l^{**}} \bigg|_{ \boldsymbol{\lambda}^{**} =  \boldsymbol{\lambda}^{*}} \Bigg ] 
    + \mathcal{O} (|| \boldsymbol{\lambda}^{**} - \boldsymbol{\lambda}^{*} ||^3) .
  \end{nalign*}
The first two terms are zero, because the Kullback-Leibler divergence
obtains a minimum value of zero at $\boldsymbol{\lambda}^{**} =  \boldsymbol{\lambda}^{*}$.
Dropping them and the third degree term, we are left with the approximation
  \begin{nalign*}
\mathcal{D}_{\text{KL}} (p (\cdot | \boldsymbol{\lambda}^{*}) || p (\cdot | \boldsymbol{\lambda}^{**})) 
  \approx \frac{1}{2} \sum_{k = 1}^{M} \sum_{l = 1}^{M} \Bigg [ ( \lambda_k^{**} - \lambda_k^{*}) ( \lambda_l^{**} - \lambda_l^{*}) 
  \frac{\partial^2 \mathcal{D}_{\text{KL}} (p  (\cdot | \boldsymbol{\lambda}^{*}) || p  (\cdot |  \boldsymbol{\lambda}^{**}) )}{\partial \lambda_k^{**} \partial \lambda_l^{**}} \bigg|_{ \boldsymbol{\lambda}^{**} =  \boldsymbol{\lambda}^{*}} \Bigg ] . 
\end{nalign*}
If the distributions $p (\cdot | \boldsymbol{\lambda}^{*})$ and $p (\cdot | \boldsymbol{\lambda}^{**})$
are predictive distributions, then the parameters $\boldsymbol{\lambda}^{*}$ and $\boldsymbol{\lambda}^{**}$ depend on the predictor value $\mathbf{x}$, i.e. $\boldsymbol{\lambda}^{**} = \boldsymbol{\lambda} (\mathbf{x}^{**})$.
When only one predictor variable, $x_d$, is varied, an infinitesimal change in the parameters can be written as
\begin{nalign*}
 \lambda_k^{**} - \lambda_k^{*} = \frac{\partial \lambda_k^{**}}{\partial x_d^{**}} (x_d^{**} - x_d^{*}) .
\end{nalign*}
Thus we get
\begin{nalign*}
& \mathcal{D}_{\text{KL}} (p (\cdot | \boldsymbol{\lambda}^{*}) || p (\cdot | \boldsymbol{\lambda}^{**})) 
  \approx \frac{1}{2} (x_d^{**} - x_d^{*})^2 
  \sum_{k = 1}^{M} \sum_{l = 1}^{M} \left ( \frac{\partial \lambda_k^{**}}{\partial x_d^{**}} \right) \left ( \frac{\partial \lambda_l^{**}}{\partial x_d^{**}} \right) \frac{\partial^2 \mathcal{D}_{\text{KL}} (p  (\cdot | \boldsymbol{\lambda}^{*}) || p  (\cdot |  \boldsymbol{\lambda}^{**}) )}{\partial \lambda_k^{**} \partial \lambda_l^{**}} \bigg|_{ \boldsymbol{\lambda}^{**} =  \boldsymbol{\lambda}^{*}}  . 
\end{nalign*}
Rearranging the terms gives the approximate equivalence
\begin{nalign*}
 \frac{2 \mathcal{D}_{\text{KL}} (p (\cdot | \boldsymbol{\lambda}^{*}) || p (\cdot | \boldsymbol{\lambda}^{**}))}{(x_d^{**} - x_d^{*})^2} 
& \approx \sum_{k = 1}^{M} \sum_{l = 1}^{M} \left ( \frac{\partial \lambda_k^{**}}{\partial x_d^{**}} \right) \left ( \frac{\partial \lambda_l^{**}}{\partial x_d^{**}} \right) \frac{\partial^2 \mathcal{D}_{\text{KL}} (p  (\cdot | \boldsymbol{\lambda}^{*}) || p  (\cdot |  \boldsymbol{\lambda}^{**}) )}{\partial \lambda_k^{**} \partial \lambda_l^{**}} \bigg|_{ \boldsymbol{\lambda}^{**} =  \boldsymbol{\lambda}^{*}} \\
& = \frac{\partial^2 \mathcal{D}_{\text{KL}} (p (\cdot | \boldsymbol{\lambda}^{*}) || p (\cdot | \boldsymbol{\lambda}^{**}))}{(\partial x_d^{**})^2}   \bigg|_{ \boldsymbol{\lambda}^{**} =  \boldsymbol{\lambda}^{*} } .
\end{nalign*}
The last identity is based on the chain rule of differentiation.
Finally, taking the square root gives the approximate equivalence
\begin{nalign*}
& \frac{\sqrt{2 \mathcal{D}_{\text{KL}} (p (\cdot | \boldsymbol{\lambda}^{*}) || p (\cdot | \boldsymbol{\lambda}^{**}))}}{|x_d^{**} - x_d^{*}|} 
 \approx \sqrt{ \frac{\partial^2 \mathcal{D}_{\text{KL}} (p (\cdot | \boldsymbol{\lambda}^{*}) || p (\cdot | \boldsymbol{\lambda}^{**}))}{(\partial x_d^{**})^2}   \bigg|_{ \boldsymbol{\lambda}^{**} =  \boldsymbol{\lambda}^{*} } },
\end{nalign*}
where the left hand side is the finite difference KL method of~\citet{paananen2019variable} and
the right hand side is the R-sens measure with $\alpha = 1$.

\clearpage

\section{R-Sens$_2$ Approximation Benefits}

\begin{figure}[h]
  \centering
    \includegraphics[width=0.8\textwidth]{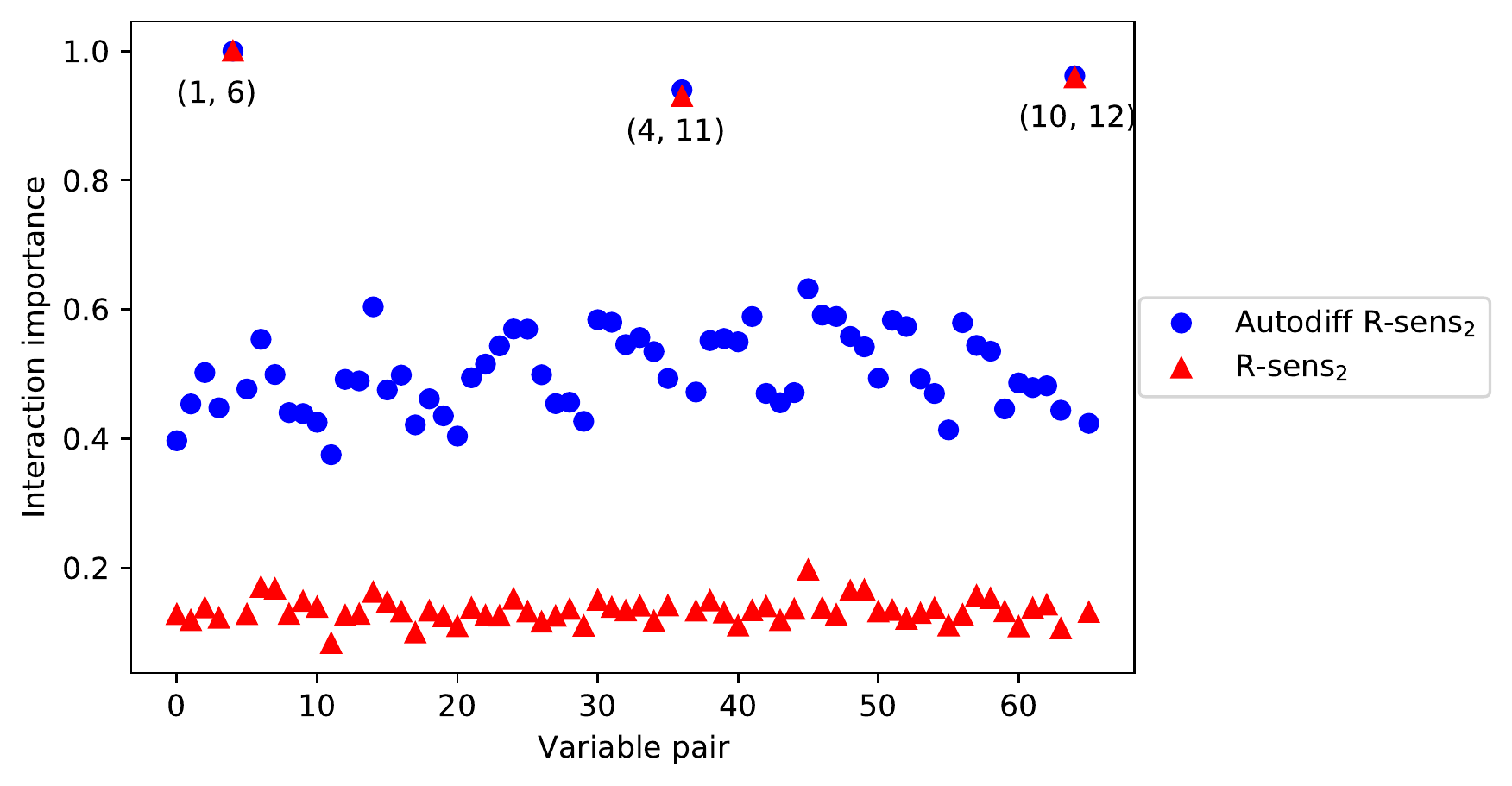}
\caption{Comparison of interaction importance estimates for R-sens$_2$ (red) and the fourth derivative of
the R\'{e}nyi divergence (blue).}\label{fig:autodiff_illustr}
\end{figure}

In this section, we show an example of how the simplified R-sens$_2$ formula we use is better
than the full formula that includes cross-derivative terms. With full formula we mean the fourth derivative of
the R\'{e}nyi divergence without dropping any terms.
We replicate the simulation experiment from
Section 3.2 of the main paper such that we compute interaction importance estimates using the simplified R-sens$_2$ formula
and the full formula that is obtained with automatic differentiation. In Figure~\ref{fig:autodiff_illustr}
we show the different interaction importances for a single simulation. The three annotated pairs
are the true simulated interactions, whereas all the other interactions are irrelevant.
The figure shows that the two formulas give almost equivalent importances for the true interactions, but
the simplified formula gives much lower importance estimates for the irrelevant interactions, thus
having significantly better ability to separate true interactions from nonexisting interactions.

\clearpage

\section{COMPUTATIONAL COST DETAILS}

Here, we discuss the computational cost of the variable importance methods used in the main paper.
Let us denote the number of observations with $N$ and the number of predictor variables with $D$.
Let us also denote the number of possible pairwise interactions with $\frac{D(D+1)}{2} \equiv D_2$.
Let us denote the costs of making predictions from a regression model with a location-scale likelihood
with 
\begin{itemize}
\centering
\item $C_{\text{E}}$ cost of $\text{E} [y]$,
\item $C_{\text{V}}$ cost of $\text{Var} [y]$,
\item $\tilde{C}_{\text{E}}$ cost of $\frac{\partial \text{E} [y]}{\partial x_d}$,
\item $\tilde{C}_{\text{V}}$ cost of $\frac{\partial \text{Var} [y]}{\partial x_d}$ ,
\item $\widehat{C}_{\text{E}}$ cost of $\frac{\partial^2 \text{E} [y]}{\partial x_d^2}$,
\item $\widehat{C}_{\text{V}}$ cost of $\frac{\partial^2 \text{Var} [y]}{\partial x_d^2}$ .
\end{itemize}
The computational cost of the variable importance methods can be tuned based on the amount of computational
resources. We tried to tune the cost of each method roughly equal in order to make the comparison fair.
The computational costs that were used in the experiment of Section 3.1
in the main paper are shown in Table~\ref{tab:compcost}, and
the costs used in the Concrete data experiment of Section 3.3
are shown in Table~\ref{tab:compcost2}.

\begin{table}[h]
\centering
\caption{Computational costs of the variable importance methods used in the first experiment of the main paper.}
\begin{tabular}[tbp]{ c c }
\toprule
Method & Time complexity  \\
\midrule
R-sens& $N D (C_{\text{V}} + \tilde{C}_{\text{E}} + \tilde{C}_{\text{V}})$  \\
EAD & $N D \tilde{C}_{\text{E}}$  \\
AED & $N D \tilde{C}_{\text{E}}$  \\
APC & $ 2 N D C_{\text{E}} + D^3$  \\
SHAP & $ 2 N D C_{\text{E}}$  \\
PD & $9 N D C_{\text{E}}$  \\
PFI & $ N (D+1) (C_{\text{E}} + C_{\text{V}})$  \\
VAR & $N D C_{\text{E}} + D^3$  \\

\bottomrule

\end{tabular}
\label{tab:compcost}
\end{table}

\begin{table}[h]
\centering
\caption{Computational costs of the variable importance methods (for interactions) used in the Concrete experiment of the main paper.}
\begin{tabular}[tbp]{ c c }
\toprule
Method & Time complexity \\
\midrule
R-sens& $N D_2 (C_{\text{V}} + \widehat{C}_{\text{E}} + \widehat{C}_{\text{V}})$  \\
EAH & $N D_2 \widehat{C}_{\text{E}}$  \\
AEH & $N D_2 \widehat{C}_{\text{E}}$  \\
SHAP & $ 4 N D_2 C_{\text{E}}$  \\
PD & $9 N D_2 C_{\text{E}}$  \\
HS & $N^2 D_2 C_{\text{E}}$  \\

\bottomrule

\end{tabular}
\label{tab:compcost2}
\end{table}

\clearpage

\section{SIMULATED INDIVIDUAL EFFECTS - ADDITIONAL RESULTS}

\subsection{Different Distributions for the Predictors}

\subsubsection{Gaussian}

\begin{table*}[tbp]
\centering
\caption{Average error in rankings in the simulated example of the main paper.
Each predictor has an independent standard normal distribution.}
\scalebox{0.9}{
\begin{tabular}{ c }
 \, \\
 \textbf{Ground-truth models}  \\
\end{tabular}
}
\scalebox{0.9}{
\begin{tabular}[tbp]{ c c c c c c c c c c }
\toprule
\multicolumn{2}{c}{Function $f_{\text{true},i} (x)$} & R-sens & EAD & AED & APC & SHAP & PD & PFI & VAR  \\
\midrule
\begin{minipage}[c]{6mm}
\centering
    \includegraphics[width=6mm]{figs/boxplot1.pdf}
\end{minipage} & $x$  &
$ \mathbf{0}$ & $\mathbf{0}$ & $\mathbf{0}$ & $\mathbf{0}$ & $0.9 \pm 0.1$ & $0.7 \pm 0.1$ & $1.0 \pm 0.1$ & \; $0.4 \pm 0.1$
 \\
\hline
\begin{minipage}[c]{6mm}
\centering
    \includegraphics[width=6mm]{figs/boxplot2.pdf}
\end{minipage} & $x^3$ &
$ \mathbf{0}$ & $\mathbf{0.0 \pm 0.1}$ &  \, $\mathbf{0.0 \pm 0.1}$ & \, $5.9 \pm 0.4$ & $2.8 \pm 0.2$ & $3.3 \pm 0.2$ & $4.2 \pm 0.3$ & \; $2.0 \pm 0.2$
\\
\hline
\begin{minipage}[c]{6mm}
\centering
    \includegraphics[width=6mm]{figs/boxplot3.pdf}
\end{minipage} &$x + \cos (3 x)$  &
$\mathbf{0}$ & $ \mathbf{0.0 \pm 0.0}$ &  \, $3.9 \pm 0.2$ & \, $8.0 \pm 0.3$ & $0.8 \pm 0.1$ & $\mathbf{0.0 \pm 0.1}$ & $0.7 \pm 0.1$ & \; $\mathbf{0.0 \pm 0.1}$
\\
\hline
\begin{minipage}[c]{6mm}
\centering
    \includegraphics[width=6mm]{figs/boxplot5.pdf}
\end{minipage} &$\sin (3 x)$  &
$ \mathbf{0}$ & $\mathbf{0.0 \pm 0.0}$ & $21.0 \pm 0.6$ & $10.8 \pm 0.3$ & $0.4 \pm 0.1$ & $0.1 \pm 0.0$ & $0.3 \pm 0.1$ & \; $3.2 \pm 0.2$
\\
\hline
\begin{minipage}[c]{6mm}
\centering
    \includegraphics[width=6mm]{figs/boxplot6.pdf}
\end{minipage} &$x \exp (-x)$  &
$ \mathbf{0}$ & $0.4 \pm 0.1$ & \, $0.5 \pm 0.1$ & \, $7.9 \pm 0.4$ & $3.1 \pm 0.3$ & $1.6 \pm 0.2$ & $6.0 \pm 0.3$ & \; $2.2 \pm 0.2$
\\
\hline
\begin{minipage}[c]{6mm}
\centering
    \includegraphics[width=6mm]{figs/boxplot7.pdf}
\end{minipage} &$ \exp (-x^2)$  &
$0$ & $0.0 \pm 0.0$ & $20.5 \pm 0.5$ & \, $7.2 \pm 0.3$ & $0.5 \pm 0.1$ & $0.4 \pm 0.1$ & $0.4 \pm 0.1$ & $ \mathbf{-0.1 \pm 0.0}$
\\
\bottomrule
\vspace{-3mm} \\
\multicolumn{10}{c}{\textbf{Imperfect models}} \\
\toprule
\multicolumn{2}{c}{Function $f_{\text{true},i} (x)$} & R-sens & EAD & AED & APC & SHAP & PD & PFI & VAR  \\

\midrule
\begin{minipage}[c]{6mm}
\centering
    \includegraphics[width=6mm]{figs/boxplot1.pdf}
\end{minipage} & $x$  &
$ \mathbf{0}$ & $0.2 \pm 0.1$ & \, $\mathbf{0.1 \pm 0.2}$ & \, $\mathbf{0.2 \pm 0.2}$ & $1.3 \pm 0.2$ & \; $\mathbf{0.1 \pm 0.1}$ & \, $1.9 \pm 0.2$ & \; $0.2 \pm 0.1$ \\
\hline
\begin{minipage}[c]{6mm}
\centering
    \includegraphics[width=6mm]{figs/boxplot2.pdf}
\end{minipage} & $x^3$&
$ \mathbf{0}$ & $\mathbf{0.2 \pm 0.3}$ & \, $\mathbf{0.2 \pm 0.3}$ & \, $6.2 \pm 0.4$ & $5.1 \pm 0.4$ & \; $2.0 \pm 0.3$ & \, $7.5 \pm 0.5$ & \; $1.1 \pm 0.3$ \\
\hline
\begin{minipage}[c]{6mm}
\centering
    \includegraphics[width=6mm]{figs/boxplot3.pdf}
\end{minipage} &$x + \cos (3 x)$  &
$\mathbf{0}$ & $\mathbf{0.0 \pm 0.1}$ & \, $4.0 \pm 0.2$ & \, $8.1 \pm 0.3$ & $1.5 \pm 0.2$ & $ \mathbf{-0.1 \pm 0.1}$ & \, $1.7 \pm 0.2$ & \; $\mathbf{0.0 \pm 0.1}$
\\
\hline
\begin{minipage}[c]{6mm}
\centering
    \includegraphics[width=6mm]{figs/boxplot5.pdf}
\end{minipage} &$\sin (3 x)$  &
$ \mathbf{0}$ & $\mathbf{0.0 \pm 0.0}$ & $20.9 \pm 0.5$ & $10.4 \pm 0.3$ & $0.4 \pm 0.1$ & \; $\mathbf{0.1 \pm 0.1}$ & \, $0.4 \pm 0.1$ & \; $3.2 \pm 0.2$ \\
\hline
\begin{minipage}[c]{6mm}
\centering
    \includegraphics[width=6mm]{figs/boxplot6.pdf}
\end{minipage} &$x \exp (-x)$  &
$ \mathbf{0}$ & $0.5 \pm 0.3$ & \, $0.7 \pm 0.3$ & \, $8.3 \pm 0.5$ & $5.8 \pm 0.4$ & \; $2.8 \pm 0.4$ & $10.0 \pm 0.5$ & \; $2.6 \pm 0.3$ \\
\hline
\begin{minipage}[c]{6mm}
\centering
    \includegraphics[width=6mm]{figs/boxplot7.pdf}
\end{minipage} &$ \exp (-x^2)$  &
$0$ & $0.0 \pm 0.0$ & $20.5 \pm 0.5$ & \, $7.2 \pm 0.3$ & $0.5 \pm 0.1$ & \; $0.5 \pm 0.1$ & \, $0.4 \pm 0.1$ & $ \mathbf{-0.1 \pm 0.0}$
\\
\bottomrule

\end{tabular}
}
\label{tab:truemaineffx2}
\end{table*}

\subsubsection{Mixture of 2 Gaussians}

\begin{table*}[h]
\centering
\caption{Average error in rankings in the simulated example of the main paper.
Each predictor is independently distributed with a mixture of 2 Gaussians.}
\scalebox{0.9}{
\begin{tabular}{ c }
 \, \\
 \textbf{Ground-truth models}  \\
\end{tabular}
}
\scalebox{0.9}{
\begin{tabular}[tbp]{ c c c c c c c c c c }
\toprule
\multicolumn{2}{c}{Function $f_{\text{true},i} (x)$} & R-sens & EAD & AED & APC & SHAP & PD & PFI & VAR  \\
\midrule
\begin{minipage}[c]{6mm}
\centering
    \includegraphics[width=6mm]{figs/boxplot1.pdf}
\end{minipage} & $x$  & 
$ \mathbf{0}$ & $\mathbf{0.0 \pm 0.0}$ & $\mathbf{0.0 \pm 0.0}$ & $\mathbf{0.0 \pm 0.0}$ & $0.5 \pm 0.1$ & $0.1 \pm 0.0$ & $0.4 \pm 0.1$ & $\mathbf{0.0 \pm 0.0}$
 \\
\hline
\begin{minipage}[c]{6mm}
\centering
    \includegraphics[width=6mm]{figs/boxplot2.pdf}
\end{minipage} & $x^3$ & 
$0$ & $ \mathbf{0.0 \pm 0.0}$ & $0.0 \pm 0.0$ & $6.6 \pm 0.4$ & $1.6 \pm 0.2$ & $1.4 \pm 0.1$ & $2.0 \pm 0.2$ & $1.0 \pm 0.1$
 \\
\hline
\begin{minipage}[c]{6mm}
\centering
    \includegraphics[width=6mm]{figs/boxplot3.pdf}
\end{minipage} &$x + \cos (3 x)$  & 
$0$ & $0.0 \pm 0.0$ & $3.8 \pm 0.2$ & $6.5 \pm 0.3$ & $0.6 \pm 0.1$ & $0.1 \pm 0.1$ & $0.5 \pm 0.1$ & $ \mathbf{-0.1 \pm 0.0}$
\\
\hline
\begin{minipage}[c]{6mm}
\centering
    \includegraphics[width=6mm]{figs/boxplot5.pdf}
\end{minipage} &$\sin (3 x)$  & 
$ \mathbf{0}$ & $\mathbf{0.0 \pm 0.0}$ & $3.6 \pm 0.2$ & $8.1 \pm 0.3$ & $0.5 \pm 0.1$ & $\mathbf{0.1 \pm 0.1}$ & $0.3 \pm 0.1$ & $2.3 \pm 0.2$
\\
\hline
\begin{minipage}[c]{6mm}
\centering
    \includegraphics[width=6mm]{figs/boxplot6.pdf}
\end{minipage} &$x \exp (-x)$  & 
$ \mathbf{0}$ & $\mathbf{0.1 \pm 0.1}$ & $0.2 \pm 0.1$ & $8.2 \pm 0.4$ & $1.2 \pm 0.2$ & $0.5 \pm 0.2$ & $1.8 \pm 0.2$ & $2.5 \pm 0.2$
\\
\hline
\begin{minipage}[c]{6mm}
\centering
    \includegraphics[width=6mm]{figs/boxplot7.pdf}
\end{minipage} &$ \exp (-x^2)$  & 
$\mathbf{0}$ & $\mathbf{0.0 \pm 0.0}$ & $20.4 \pm 0.5$ & $9.8 \pm 0.4$ & $0.7 \pm 0.1$ & $0.3 \pm 0.1$ & $0.6 \pm 0.1$ & $ \mathbf{0.0 \pm 0.0}$
  \\
\bottomrule
\multicolumn{10}{c}{\textbf{Imperfect models}} \\
\toprule
\multicolumn{2}{c}{Function $f_{\text{true},i} (x)$} & R-sens & EAD & AED & APC & SHAP & PD & PFI & VAR  \\

\midrule
\begin{minipage}[c]{6mm}
\centering
    \includegraphics[width=6mm]{figs/boxplot1.pdf}
\end{minipage} & $x$  & 
$0$ & $0.1 \pm 0.2$ & $0.2 \pm 0.2$ & $-0.6 \pm 0.1$ & $-0.0 \pm 0.2$ & $ \mathbf{-0.8 \pm 0.1}$ & $-0.2 \pm 0.1$ & $-0.7 \pm 0.1$
\\
\hline
\begin{minipage}[c]{6mm}
\centering
    \includegraphics[width=6mm]{figs/boxplot2.pdf}
\end{minipage} & $x^3$&
$ \mathbf{0}$ & $\mathbf{0.1 \pm 0.2}$ & $\mathbf{0.1 \pm 0.2}$ & $6.0 \pm 0.4$ & $1.4 \pm 0.2$ & $0.4 \pm 0.2$ & $2.5 \pm 0.2$ & $1.3 \pm 0.2$
 \\
\hline
\begin{minipage}[c]{6mm}
\centering
    \includegraphics[width=6mm]{figs/boxplot3.pdf}
\end{minipage} &$x + \cos (3 x)$  & 
$0$ & $0.0 \pm 0.1$ & $3.9 \pm 0.2$ & $6.6 \pm 0.3$ & $0.7 \pm 0.1$ & $-0.1 \pm 0.1$ & $0.7 \pm 0.1$ & $ \mathbf{-0.3 \pm 0.1}$
\\
\hline
\begin{minipage}[c]{6mm}
\centering
    \includegraphics[width=6mm]{figs/boxplot5.pdf}
\end{minipage} &$\sin (3 x)$  & 
$\mathbf{0}$ & $ \mathbf{0.0 \pm 0.0}$ & $3.7 \pm 0.2$ & $7.7 \pm 0.3$ & $0.5 \pm 0.1$ & $\mathbf{0.1 \pm 0.1}$ & $0.4 \pm 0.1$ & $1.9 \pm 0.2$
\\
\hline
\begin{minipage}[c]{6mm}
\centering
    \includegraphics[width=6mm]{figs/boxplot6.pdf}
\end{minipage} &$x \exp (-x)$  & 
$\mathbf{0}$ & $\mathbf{0.1 \pm 0.2}$ & $\mathbf{0.2 \pm 0.2}$ & $7.8 \pm 0.4$ & $1.1 \pm 0.3$ & $ \mathbf{0.0 \pm 0.2}$ & $2.6 \pm 0.3$ & $3.9 \pm 0.3$
\\
\hline
\begin{minipage}[c]{6mm}
\centering
    \includegraphics[width=6mm]{figs/boxplot7.pdf}
\end{minipage} &$ \exp (-x^2)$  & 
$\mathbf{0}$ & $\mathbf{0.0 \pm 0.0}$ & $20.4 \pm 0.5$ & $9.3 \pm 0.4$ & $0.7 \pm 0.1$ & $0.3 \pm 0.1$ & $0.6 \pm 0.1$ & $ \mathbf{0.0 \pm 0.0}$
\\
\bottomrule

\end{tabular}
}
\label{tab:maineffx3}
\end{table*}

\subsubsection{Correlated Gaussian}

\begin{table*}[h]
\centering
\caption{Average error in rankings in the simulated example of the main paper.
The predictors have a multivariate Normal distribution with all correlations 0.8.}
\scalebox{0.9}{
\begin{tabular}{ c }
 \, \\
 \textbf{Ground-truth models}  \\
\end{tabular}
}
\scalebox{0.9}{
\begin{tabular}[tbp]{ c c c c c c c c c c }
\toprule
\multicolumn{2}{c}{Function $f_{\text{true},i} (x)$} & R-sens & EAD & AED & APC & SHAP & PD & PFI & VAR  \\
\midrule
\begin{minipage}[c]{6mm}
\centering
    \includegraphics[width=6mm]{figs/boxplot1.pdf}
\end{minipage} & $x$  & 
$ \mathbf{0}$ & $\mathbf{0.0 \pm 0.0}$ & $\mathbf{0.0 \pm 0.0}$ & $\mathbf{0.0 \pm 0.0}$ & $0.5 \pm 0.1$ & $0.1 \pm 0.0$ & $3.5 \pm 0.2$ & $0.4 \pm 0.1$
 \\
\hline
\begin{minipage}[c]{6mm}
\centering
    \includegraphics[width=6mm]{figs/boxplot2.pdf}
\end{minipage} & $x^3$ & 
$ \mathbf{0}$ & $\mathbf{0.0 \pm 0.0}$ & $\mathbf{0.0 \pm 0.0}$ & $4.6 \pm 0.4$ & $2.6 \pm 0.2$ & $2.8 \pm 0.2$ & $5.2 \pm 0.3$ & $1.1 \pm 0.1$
 \\
\hline
\begin{minipage}[c]{6mm}
\centering
    \includegraphics[width=6mm]{figs/boxplot3.pdf}
\end{minipage} &$x + \cos (3 x)$  & 
$\mathbf{0}$ & $\mathbf{0.0 \pm 0.0}$ & $3.4 \pm 0.2$ & $6.1 \pm 0.3$ & $0.5 \pm 0.1$ & $\mathbf{-0.1 \pm 0.1}$ & $2.5 \pm 0.2$ & $ \mathbf{-0.1 \pm 0.1}$

\\
\hline
\begin{minipage}[c]{6mm}
\centering
    \includegraphics[width=6mm]{figs/boxplot5.pdf}
\end{minipage} &$\sin (3 x)$  & 
$\mathbf{0}$ & $\mathbf{0.0 \pm 0.0}$ & $20.8 \pm 0.5$ & $7.4 \pm 0.3$ & $0.5 \pm 0.1$ & $0.1 \pm 0.0$ & $1.0 \pm 0.1$ & $ \mathbf{0.0 \pm 0.0}$
\\
\hline
\begin{minipage}[c]{6mm}
\centering
    \includegraphics[width=6mm]{figs/boxplot6.pdf}
\end{minipage} &$x \exp (-x)$  & 
$ \mathbf{0}$ & $0.2 \pm 0.1$ & $0.3 \pm 0.1$ & $4.4 \pm 0.3$ & $3.1 \pm 0.2$ & $1.6 \pm 0.2$ & $6.7 \pm 0.3$ & $1.4 \pm 0.2$
\\
\hline
\begin{minipage}[c]{6mm}
\centering
    \includegraphics[width=6mm]{figs/boxplot7.pdf}
\end{minipage} &$ \exp (-x^2)$  & 
$ \mathbf{0}$ & $\mathbf{0.0 \pm 0.0}$ & $18.9 \pm 0.6$ & $4.6 \pm 0.3$ & $0.4 \pm 0.1$ & $\mathbf{0.1 \pm 0.1}$ & $2.1 \pm 0.2$ & $\mathbf{0.0 \pm 0.0}$
  \\
\bottomrule
\multicolumn{10}{c}{\textbf{Imperfect models}} \\
\toprule
\multicolumn{2}{c}{Function $f_{\text{true},i} (x)$} & R-sens & EAD & AED & APC & SHAP & PD & PFI & VAR  \\

\midrule
\begin{minipage}[c]{6mm}
\centering
    \includegraphics[width=6mm]{figs/boxplot1.pdf}
\end{minipage} & $x$  & 
$0$ & $0.2 \pm 0.1$ & $0.2 \pm 0.1$ & $1.3 \pm 0.2$ & $0.7 \pm 0.2$ & $ \mathbf{-0.4 \pm 0.1}$ & $4.3 \pm 0.3$ & $-0.0 \pm 0.1$
\\
\hline
\begin{minipage}[c]{6mm}
\centering
    \includegraphics[width=6mm]{figs/boxplot2.pdf}
\end{minipage} & $x^3$&
$ \mathbf{0}$ & $\mathbf{0.1 \pm 0.3}$ & $\mathbf{0.1 \pm 0.3}$ & $6.4 \pm 0.4$ & $3.8 \pm 0.4$ & $1.4 \pm 0.3$ & $9.5 \pm 0.8$ & $0.5 \pm 0.3$
 \\
\hline
\begin{minipage}[c]{6mm}
\centering
    \includegraphics[width=6mm]{figs/boxplot3.pdf}
\end{minipage} &$x + \cos (3 x)$  & 
$0$ & $0.0 \pm 0.1$ & $3.4 \pm 0.2$ & $6.6 \pm 0.3$ & $1.0 \pm 0.1$ & $-0.2 \pm 0.1$ & $3.2 \pm 0.2$ & $ \mathbf{-0.3 \pm 0.1}$
\\
\hline
\begin{minipage}[c]{6mm}
\centering
    \includegraphics[width=6mm]{figs/boxplot5.pdf}
\end{minipage} &$\sin (3 x)$  & 
$0$ & $0.0 \pm 0.0$ & $20.7 \pm 0.5$ & $7.5 \pm 0.3$ & $0.5 \pm 0.1$ & $0.0 \pm 0.1$ & $1.0 \pm 0.1$ & $ \mathbf{-0.1 \pm 0.0}$
\\
\hline
\begin{minipage}[c]{6mm}
\centering
    \includegraphics[width=6mm]{figs/boxplot6.pdf}
\end{minipage} &$x \exp (-x)$  & 
$ \mathbf{0}$ & $0.4 \pm 0.3$ & $0.5 \pm 0.3$ & $5.7 \pm 0.4$ & $4.8 \pm 0.4$ & $3.2 \pm 0.5$ & $11.1 \pm 0.8$ & $1.9 \pm 0.3$
\\
\hline
\begin{minipage}[c]{6mm}
\centering
    \includegraphics[width=6mm]{figs/boxplot7.pdf}
\end{minipage} &$ \exp (-x^2)$  & 
$ \mathbf{0}$ & $\mathbf{0.0 \pm 0.0}$ & $18.9 \pm 0.6$ & $4.6 \pm 0.3$ & $0.5 \pm 0.1$ & $\mathbf{0.1 \pm 0.1}$ & $2.1 \pm 0.2$ & $\mathbf{0.0 \pm 0.0}$
\\
\bottomrule

\end{tabular}
}
\label{tab:maineffx4}
\end{table*}

\clearpage

\section{R-SENS FOR GAUSSIAN PROCESS MODELS}

Gaussian process models are widely used in supervised learning, where
the task is to predict an output $y$ from a $D$-dimensional input
$\mathbf{x}$.
The type of functions the Gaussian process can represent
are determined by its covariance function, which is a key decision made during modelling.
The covariance function $k (\mathbf{x}^{(i)},\mathbf{x}^{(j)})$ defines
the covariance between the function values at the input points $\mathbf{x}^{(i)}$ and $\mathbf{x}^{(j)}$.
We assume the Gaussian process has a zero mean, in which case the joint distribution
of the latent output values $f$ at the training points is
\begin{equation*}
p ( f (\mathbf{X})) = p ( \mathbf{f}) = \text{Normal} (\mathbf{f} \, | \, 0 , \mathbf{K}) ,
\end{equation*}
where $\mathbf{K}$ is the covariance matrix between the latent function values
at the training inputs $\mathbf{X} = (\mathbf{x}^{(1)}, \ldots , \mathbf{x}^{(N)})$
such that
$\mathbf{K}_{ij} = k (\mathbf{x}^{(i)},\mathbf{x}^{(j)})$.

In this work, we use the exponentiated quadratic covariance function
\begin{nalign} \label{eq:ARD-SE}
& k_{\mathrm{EQ}} (\mathbf{x}^{(i)},\mathbf{x}^{(j)}) = 
 \sigma_f^2 \, \mathrm{exp} \left( - \frac{1}{2} \sum_{k = 1}^D \frac{(x_k^{(i)} - x_k^{(j)})^2}{l_k^2} \right ).
\end{nalign}
Here, the hyperparameter~$\sigma_f$ determines the overall variability of the functions, and $( l_1 , ... , l_D )$ are the length-scales
of each input dimension.
By defining an observation model that links the observations to the latent
values of the Gaussian process, the model can be used for inference and predictions in many supervised learning tasks.

For example, in regression with an assumption of Gaussian noise, the posterior distribution of latent values
for a new input point $\mathbf{x}^*$ is a univariate normal distribution with mean and variance
\begin{nalign} \label{eq:gauss_pred}
& \text{E}   [f^{*} | \mathbf{x}^{*}, \mathbf{y}] = k (\mathbf{x}^{*}, \mathbf{X}) (k(\mathbf{X},\mathbf{X}) + \sigma^2 \mathbf{I})^{-1} \mathbf{y} \\
& \text{Var} [f^{*} | \mathbf{x}^{*}, \mathbf{y}] = k (\mathbf{x}^{*}, \mathbf{x}^{*})  -  k (\mathbf{x}^{*}, \mathbf{X}) 
(k(\mathbf{X},\mathbf{X}) + \sigma^2 \mathbf{I})^{-1}  k (\mathbf{X}, \mathbf{x}^{*}),
\end{nalign}
where $\sigma^2$ is the noise variance, $\mathbf{I}$ is the identity matrix, and $\mathbf{y}$ is the vector of training outputs.
For many other observation models, the posterior of latent values is not Gaussian, but
it is commonplace to approximate it with a Gaussian distribution during inference, and
many methods
have been developed for doing so~\citep{williams1998bayesian,Opper+Winther:2000,minka2001family,rasmussen2006gaussian}.
The variable importance assessment thus depends implicitly on
the posterior approximation, as does any general method that uses the model's predictions.

\subsection{Differentiating Gaussian Processes}

We assume that the posterior distribution of latent values is Gaussian.
Because differentiation is a linear operation, the derivatives
of the parameters of a Gaussian process posterior distribution with respect to predictor variables are available in closed form~\citep{solak2003derivative,rasmussen2003gaussian}.
For example, for the Gaussian observation model, the derivatives of the mean and variance of the predictive distribution in equation~(\ref{eq:gauss_pred}) with respect to the predictor variable $x_d$ at point $\mathbf{x}^{*}$ are given as
\begin{nalign*}
 \frac{\partial \text{E}   [f^{*} | \mathbf{x}^{*}, \mathbf{y}] }{\partial x^{*}_d} & = \frac{\partial k (\mathbf{x}^{*}, \mathbf{X}) }{\partial x^{*}_d} (k(\mathbf{X},\mathbf{X}) + \sigma^2 \mathbf{I})^{-1} \mathbf{y} \\
 \frac{\partial \text{Var}  [f^{*} | \mathbf{x}^{*}, \mathbf{y}] }{\partial x^{*}_d} & = \frac{\partial k (\mathbf{x}^{*}, \mathbf{x}^{*}) }{\partial x^{*}_d} 
  -   \frac{\partial k (\mathbf{x}^{*}, \mathbf{X}) }{\partial x^{*}_d}
 (k(\mathbf{X},\mathbf{X}) + \sigma^2 \mathbf{I})^{-1} k (\mathbf{X}, \mathbf{x}^{*}) 
  -    k (\mathbf{x}^{*}, \mathbf{X})
 (k(\mathbf{X},\mathbf{X}) + \sigma^2 \mathbf{I})^{-1} \frac{\partial k (\mathbf{X}, \mathbf{x}^{*}) }{\partial x^{*}_d} .
\end{nalign*}
For the exponentiated quadratic covariance function in equation~(\ref{eq:ARD-SE}), the partial
derivatives with respect to the predictor variable $x_d$ are
\begin{nalign*}
& \frac{\partial k_{\text{EQ}} (\mathbf{x}^{(i)},\mathbf{x}^{(j)}) }{\partial x^{(i)}_d} =  
 \sigma_f^2 \, \mathrm{exp} \left( - \frac{1}{2} \sum_{k = 1}^D \frac{(x_k^{(i)} - x_k^{(j)})^2}{l_k^2} \right ) \left (- \frac{x_d^{(i)} - x_d^{(j)}}{l_d^2} \right ) ,\\
& \frac{\partial k_{\text{EQ}} (\mathbf{x}^{(i)},\mathbf{x}^{(j)}) }{\partial x^{(j)}_d} = - \frac{\partial k_{\text{EQ}} (\mathbf{x}^{(i)},\mathbf{x}^{(j)}) }{\partial x^{(i)}_d} .
\end{nalign*}
The second derivatives are
\begin{nalign*}
 \frac{\partial^2 k_{\text{EQ}} (\mathbf{x}^{(i)},\mathbf{x}^{(j)}) }{\partial x^{(i)}_d \partial x^{(i)}_e} = \frac{\partial^2 k_{\text{EQ}} (\mathbf{x}^{(i)},\mathbf{x}^{(j)}) }{\partial x^{(j)}_d \partial x^{(j)}_e} = 
 \sigma_f^2 \, \mathrm{exp} \left( -\frac{1}{2} \sum_{k = 1}^D \frac{(x_k^{(i)} - x_k^{(j)})^2}{l_k^2} \right ) 
\left ( \frac{x_d^{(i)} - x_d^{(j)}}{l_d^2} \right ) \left ( \frac{x_e^{(i)} - x_e^{(j)}}{l_e^2} \right ) .
\end{nalign*}

For the R-sens measure, we need derivatives with respect to the
parameters of the predictive distribution and not the posterior of the latent values.
However, for many observation models these are obtained as a function of the derivatives
of the latent values.
In this section, we derive
the equations
for some commonly used observation models.

\subsection{Regression with Gaussian Observation Model}

In regression problems, it is commonly assumed that
the noise has a Gaussian distribution.
For a Gaussian observation model, the predictive distribution for a new observation $y^{*}$ at a single predictor value
$\mathbf{x}^{*}$ is a normal distribution, which we will denote
\begin{nalign*}
p (y^{*} | \mathbf{x}^{*}, \mathbf{y}) = \text{Normal} (y^{*} | \text{E} [y^{*}], \text{Var} [y^{*}])
= \text{Normal} (y^{*} | \text{E} [f^{*}], \text{Var} [f^{*}] + \sigma^2) ,
\end{nalign*}
where $\text{E} [f^{*}]$ and $\text{Var} [f^{*}]$ are the mean and variance of the posterior distribution of
latent values at $\mathbf{x}^{*}$, and $\sigma^2$ is the noise variance.
Now, the derivatives of $\text{E} [y^{*}]$ and $\text{Var} [y^{*}]$ with respect to predictor variable $x_d^{*}$ are simply
\begin{nalign*}
&  \frac{\partial \text{E} [y^{*}] }{\partial x_d^{*}} =  \frac{\partial \text{E} [f^{*}] }{\partial x_d^{*}} \\
&  \frac{\partial \text{Var} [y^{*}] }{\partial x_d^{*}} =  \frac{\partial \text{Var} [f^{*}] }{\partial x_d^{*}} \\
&  \frac{\partial^2 \text{E} [y^{*}] }{\partial x_d^{*} \partial x_e^{*}} =  \frac{\partial^2 \text{E} [f^{*}] }{\partial x_d^{*} \partial x_e^{*}} \\
&  \frac{\partial^2 \text{Var} [y^{*}] }{\partial x_d^{*} \partial x_e^{*}} =  \frac{\partial^2 \text{Var} [f^{*}] }{\partial x_d^{*} \partial x_e^{*}} .
\end{nalign*}

The Fisher information elements
of the normal distribution are
\begin{nalign*}
\mathcal{I}_{\text{N}} (\text{E} [y^{*}]) & = \frac{1}{\text{Var} [y^{*}]} \\
\mathcal{I}_{\text{N}} (\text{Var} [y^{*}]) & =  \frac{1}{2 (\text{Var} [y^{*}])^2} .
\end{nalign*}
Thus, the R-sens measure takes the form
\begin{nalign*}
& \text{R-sens} (\mathbf{x}^{*},x_d, \alpha = 1)
 = 
 \sqrt{  \frac{1}{\text{Var} [y^{*}]} \left ( \frac{\partial \text{E} [f^{*}] }{\partial x_d^{*}} \right )^2 + \frac{1}{2 (\text{Var} [y^{*}])^2} \left ( \frac{\partial \text{Var} [f^{*}] }{\partial x_d^{*}} \right )^2 }.
\end{nalign*}
Here, the first term is proportional to the slope of the mean prediction scaled by the predictive
uncertainty, as
with the linear regression model discussed in Section~2.1 of the main paper.
The R-sens$_2$ measure evaluates to
\begin{nalign*}
 \text{R-sens}_2 (\mathbf{x}^{*},(x_d,x_e), \alpha = 1) 
= \sqrt{ \frac{1}{\text{Var} [y^{*}]}  \left ( \frac{\partial^2 \text{E} [f^{*}] }{\partial x_d^{*} \partial x_e^{*}} \right )^2 + \frac{1}{2 (\text{Var} [y^{*}])^2}  \left ( \frac{\partial^2 \text{Var} [f^{*}] }{\partial x_d^{*} \partial x_d^{*}} \right )^2 }.
\end{nalign*}

\subsection{Binary Classification}

For binary classification problems, the predictive distribution is a Bernoulli distribution with only one parameter, the probability of positive classification.
This is obtained by squashing the latent Gaussian process function through a link function and integrating
over the posterior of the latent function values.
Two commonly used link functions for Gaussian process
classification are the logit and probit.
The Probit link function has the benefit that the predictive
distribution has an analytical formula when the posterior distribution of latent values is approximated with a Gaussian.
Using a Probit link function, the predictive probability
has thus an approximate analytical form
\begin{nalign*}
\pi^{*} = p (y = 1 | \mathbf{x}^{*}, \mathbf{y}) = \Phi \left ( \frac{\text{E} [f^{*}]}{\sqrt{1 + \text{Var} [f^{*}]}}  \right ),
\end{nalign*}
where $\Phi$ is the cumulative distribution of the standard normal distribution.

Now, the derivatives of $\pi^{*}$ with respect to $x_d^{*}$ are
\begin{nalign*}
& \frac{\partial \pi^{*}}{\partial x_d^{*}} = \text{Normal} \left (\frac{\text{E} [f^{*}]}{\sqrt{1 + \text{Var} [f^{*}]}} \right ) 
 \left [ \frac{1}{\sqrt{1 + \text{Var} [f^{*}]}} \frac{\partial \text{E} [f^{*}]}{\partial x_d^{*}} - \frac{\text{E} [f^{*}]}{2 (1 + \text{Var} [f^{*}])^{3/2}} \frac{\partial \text{Var} [f^{*}]}{\partial x_d^{*}} \right ] ,\\
& \frac{\partial^2 \pi^{*}}{\partial x_d^{*} \partial x_e^{*}} = \text{Normal} \left (\frac{\text{E} [f^{*}]}{\sqrt{1 + \text{Var} [f^{*}]}} \right ) \left (\frac{\text{E} [f^{*}]}{\sqrt{1 + \text{Var} [f^{*}]}} \right )
 \left [ \frac{1}{\sqrt{1 + \text{Var} [f^{*}]}} \frac{\partial \text{E} [f^{*}]}{\partial x_d^{*}} - \frac{\text{E} [f^{*}]}{2 (1 + \text{Var} [f^{*}])^{3/2}} \frac{\partial \text{Var} [f^{*}]}{\partial x_d^{*}} \right ] \times \\
& \left [ \frac{1}{\sqrt{1 + \text{Var} [f^{*}]}} \frac{\partial \text{E} [f^{*}]}{\partial x_e^{*}} - \frac{\text{E} [f^{*}]}{2 (1 + \text{Var} [f^{*}])^{3/2}} \frac{\partial \text{Var} [f^{*}]}{\partial x_e^{*}} \right ] +\\
& \text{Normal} \left (\frac{\text{E} [f^{*}]}{\sqrt{1 + \text{Var} [f^{*}]}} \right ) 
 \Bigg [ \frac{1}{\sqrt{1 + \text{Var} [f^{*}]}} \frac{\partial^2 \text{E} [f^{*}]}{\partial x_d^{*} \partial x_e^{*}} - \frac{\partial \text{E} [f^{*}]}{\partial x_d^{*}} \frac{1}{2 (1 + \text{Var} [f^{*}])^{3/2}} \frac{\partial \text{Var} [f^{*}]}{\partial x_e^{*}} \\
& - \frac{\partial^2 \text{Var} [f^{*}]}{\partial x_d^{*} \partial x_e^{*}} \frac{\text{E} [f^{*}]}{2 (1 + \text{Var} [f^{*}])^{3/2}} 
 - \frac{\partial \text{Var} [f^{*}]}{\partial x_d^{*} } \Bigg ( \frac{\partial \text{E} [f^{*}]}{\partial x_e^{*} } \frac{1}{2 (1 + \text{Var} [f^{*}])^{3/2}} 
 -  \frac{3 \text{E} [f^{*}]}{4 (1 + \text{Var} [f^{*}])^{3/2}} \frac{\partial \text{Var} [f^{*}]}{\partial x_e^{*}} \Bigg ) \Bigg ] .
\end{nalign*}

The Fisher information
of the Bernoulli distribution is
\begin{nalign*}
 \mathcal{I}_{\text{Bern}} (\pi^{*}) =
\frac{1}{\pi^{*}(1 - \pi^{*})} = 
 \left (\Phi \left ( \frac{\text{E} [f^{*}]}{\sqrt{1 + \text{Var} [f^{*}]}}  \right ) \right )^{-1} \left (1 - \Phi \left ( \frac{\text{E} [f^{*}]}{\sqrt{1 + \text{Var} [f^{*}]}}  \right ) \right )^{-1} .
\end{nalign*}
The R-sens and R-sens$_2$ measures take the form
\begin{nalign*}
& \text{R-sens} (\mathbf{x}^{*},x_d, \alpha = 1)  %
 = \sqrt{ \mathcal{I}_{\text{Bern}} (\pi^{*}) \left (\frac{\partial \pi^{*}}{\partial x_d^{*}} \right )^2 }, \\
 & \text{R-sens}_2 (\mathbf{x}^{*},(x_d,x_e), \alpha = 1)  %
 = \sqrt{ \mathcal{I}_{\text{Bern}} (\pi^{*}) \left (\frac{\partial^2 \pi^{*}}{\partial x_d^{*} \partial x_e^{*}} \right )^2 }.
\end{nalign*}

\subsection{Poisson Observation Model}

For modelling count data with Gaussian processes, it is common to use
a combination of a count observation model with a link function that transforms the positively constrained parameters to unconstrained scale where the Gaussian Process prior is placed.
Here, we derive the equations needed for the R-sens method for the case of Poisson likelihood and
exponential link function.

The likelihood is
\begin{nalign*}
& p (\mathbf{y} | \mathbf{f}) = \prod _{i=1}^{n} \text{Poisson} (y_i | \lambda_i (f_i)) = \prod _{i=1}^{n} \text{Poisson} (y_i | \exp (f_i)) .
\end{nalign*}
Now, the Gaussian process prior is placed on the unconstrained latent values. If one uses a Gaussian
approximation to the posterior of the latent values, then the transformed
$\lambda$'s have a log-normal distribution.
The intensity $\lambda$ at any input point is given by integrating over
the approximate posterior $q (f^{*} | \mathbf{y}, \mathbf{x}^{*})$
\begin{nalign*}
\lambda^{*} = \int \exp (f^{*}) q (f^{*} | \mathbf{y}, \mathbf{x}^{*}) \mathrm{d} f^{*} .
\end{nalign*}
This evaluates to the mean of the log-normal distribution
\begin{nalign*}
\lambda^{*} & = \text{E} [\text{Lognormal}( \text{E} [f^{*}], \text{Var} [f^{*}])] 
 = \exp (\text{E} [f^{*}] + \text{Var} [f^{*}]/2) .
\end{nalign*}

The derivatives of this with respect to the predictor variables are
\begin{nalign*}
& \frac{\partial \lambda^{*}}{\partial x_d^{*}} =  \exp (\text{E} [f^{*}] + \text{Var} [f^{*}]/2) 
 \left ( \frac{\partial \text{E} [f^{*}] }{\partial x_d^{*}} + \frac{1}{2} \frac{\partial \text{Var} [f^{*}] }{\partial x_d^{*}} \right ) , \\
& \frac{\partial^2 \lambda^{*}}{\partial x_d^{*} \partial x_e^{*}} =  \exp (\text{E} [f^{*}] + \text{Var} [f^{*}]/2) \times \\
& \bigg [ \left ( \frac{\partial \text{E} [f^{*}] }{\partial x_d^{*}} + \frac{1}{2} \frac{\partial \text{Var} [f^{*}] }{\partial x_d^{*}} \right ) \left ( \frac{\partial \text{E} [f^{*}] }{\partial x_e^{*}} + \frac{1}{2} \frac{\partial \text{Var} [f^{*}] }{\partial x_e^{*}} \right ) + 
\left ( \frac{\partial^2 \text{E} [f^{*}] }{\partial x_d^{*} \partial x_e^{*}} + \frac{1}{2} \frac{\partial^2 \text{Var} [f^{*}] }{\partial x_d^{*} \partial x_e^{*}} \right ) \bigg ] .
\end{nalign*}

The Fisher information
of the Poisson distribution is
\begin{nalign*}
\mathcal{I}_{\text{Pois}} (\lambda^{*}) =
\frac{1}{\lambda^{*}} = 
\frac{1}{\exp (\text{E} [f^{*}] + \text{Var} [f^{*}]/2)} .
\end{nalign*}
Thus, the R-sens and R-sens$_2$ measures take the form
\begin{nalign*}
& \text{R-sens} (\mathbf{x}^{*},x_d, \alpha = 1) %
 = \sqrt{ \exp \left (\text{E} [f^{*}] + \frac{\text{Var} [f^{*}]}{2} \right ) } \left | \frac{\partial \text{E} [f^{*}] }{\partial x_d^{*}} + \frac{1}{2} \frac{\partial \text{Var} [f^{*}] }{\partial x_d^{*}} \right |, \\
& \text{R-sens}_2 (\mathbf{x}^{*},(x_d,x_e), \alpha = 1) \\
& = \sqrt{ \exp \left (\text{E} [f^{*}] + \frac{\text{Var} [f^{*}]}{2} \right ) } 
 \Bigg |
\left ( \frac{\partial \text{E} [f^{*}] }{\partial x_d^{*}} + \frac{1}{2} \frac{\partial \text{Var} [f^{*}] }{\partial x_d^{*}} \right ) \left ( \frac{\partial \text{E} [f^{*}] }{\partial x_e^{*}} + \frac{1}{2} \frac{\partial \text{Var} [f^{*}] }{\partial x_e^{*}} \right ) +
 \left ( \frac{\partial^2 \text{E} [f^{*}] }{\partial x_d^{*} \partial x_e^{*}} + \frac{1}{2} \frac{\partial^2 \text{Var} [f^{*}] }{\partial x_d^{*} \partial x_e^{*}} \right ) \Bigg |.
\end{nalign*}

\clearpage

\section{ILLUSTRATIVE EXAMPLE - LOGISTIC REGRESSION}

In this section, we show an illustrative example similar to the main paper, but with a logistic regression model
where the target variable $y$ is binary.
As the inverse link function, we use the cumulative Normal distribution.
We consider a simple multivariate Gaussian prior on the regression coefficients.
Contrary to the linear regression example, the posterior distribution has no closed form.
We will use the Laplace approximation to get
a Gaussian approximation to the posterior.
The approximate posterior is
\begin{nalign*}
p( \boldsymbol{\beta} | \mathbf{X}, \mathbf{y} ) \approx \mathcal{N} (\boldsymbol{\widehat{\beta}}, \mathbf{H}^{-1}) ,
\end{nalign*}
where $\boldsymbol{\widehat{\beta}}$ is the maximum a posteriori estimate of the regression coefficients.
When using the inverse cumulative Normal distribution as the link function, the predictive
distribution at a new point $\mathbf{x}^{*}$ has a closed form equation.
\begin{nalign*}
p(y^{*} = 1 | \mathbf{x}^{*}, \mathbf{X}, \mathbf{y}) = \pi^{*} (\mathbf{x}^{*}) = \Phi \left ( \frac{\mathbf{x}^{*} \boldsymbol{\widehat{\beta}}}{\sqrt{1 + \mathbf{x}^{*} (\mathbf{H})^{-1} \mathbf{x}^{* T}}} \right ) .
\end{nalign*}
The derivative of the success probability is
\begin{nalign*}
\frac{\partial \pi^{*} (\mathbf{x}^{*})}{\partial x_d^{*}} = \mathcal{N} \left ( \frac{\mathbf{x}^{*} \boldsymbol{\widehat{\beta}}}{\sqrt{1 + \mathbf{x}^{*} (\mathbf{H})^{-1} \mathbf{x}^{* T}}} \right ) \left [
\frac{\widehat{\beta}_d}{\sqrt{1 + \mathbf{x}^{*} (\mathbf{H})^{-1} \mathbf{x}^{* T}}} -
\frac{\mathbf{x}^{*} \boldsymbol{\widehat{\beta}} [ (\mathbf{H})^{-1} \mathbf{x}^{* T} ]_d }{(1 + \mathbf{x}^{*} (\mathbf{H})^{-1} \mathbf{x}^{* T})^{3/2}}
\right ] .
\end{nalign*}
The Fisher information of the Bernoulli distribution is
\begin{nalign*}
\mathcal{I}_{\text{Ber}} (\pi^{*}) = \frac{1}{\pi^{*} (1 - \pi^{*})} .
\end{nalign*}

The R-sens sensitivity measure for variable $x^{*}_d$ thus evaluates to
\begin{nalign} \label{eq:Rsenslog}
\sqrt{\frac{1}{\pi^{*} (1 - \pi^{*})} } \mathcal{N} \left ( \frac{\mathbf{x}^{*} \boldsymbol{\widehat{\beta}}}{\sqrt{1 + \mathbf{x}^{*} (\mathbf{H})^{-1} \mathbf{x}^{* T}}} \right )
\left |
\frac{\widehat{\beta}_d}{\sqrt{1 + \mathbf{x}^{*} (\mathbf{H})^{-1} \mathbf{x}^{* T}}} -
\frac{\mathbf{x}^{*} \boldsymbol{\widehat{\beta}} [ (\mathbf{H})^{-1} \mathbf{x}^{* T} ]_d }{(1 + \mathbf{x}^{*} (\mathbf{H})^{-1} \mathbf{x}^{* T})^{3/2}}
\right |
.
\end{nalign}

\begin{figure}[h]
  \centering
    \includegraphics[width=0.5\textwidth]{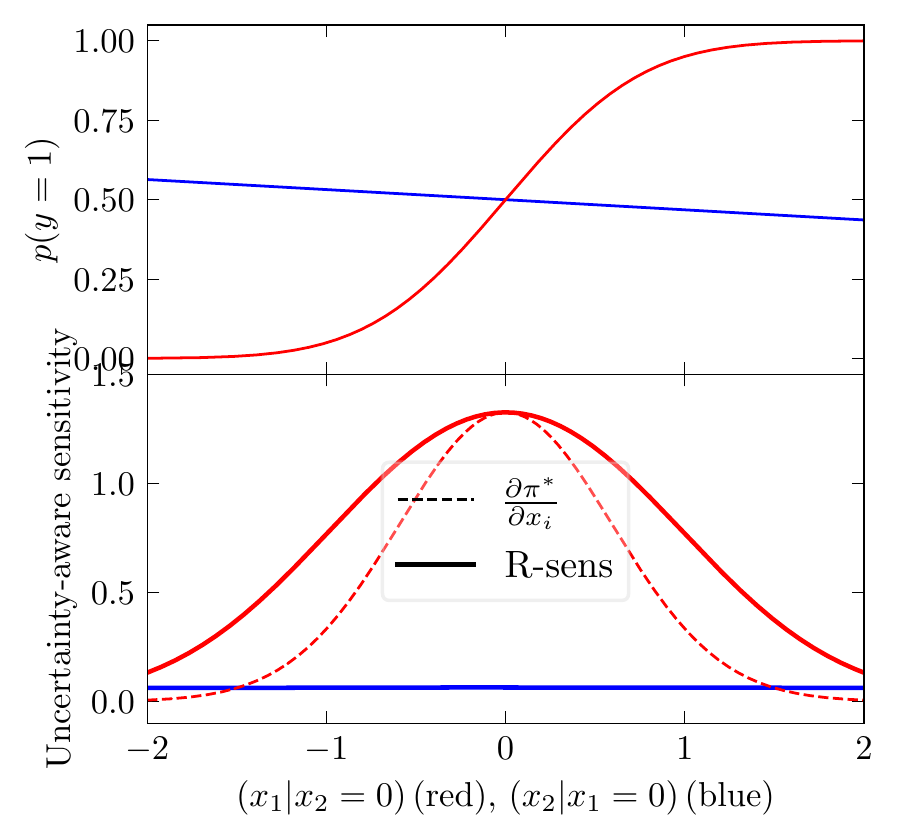}
\caption{Top: Predictive distributions $p(y = 1| x_1, x_2 = 0)$ (red) and $p(y | x_2, x_1 = 0)$ (blue) for the logistic regression model.
Bottom: R-sens uncertainty-aware sensitivity measure given by equation~(\ref{eq:Rsenslog}) for
$x_1$ (red) and $x_2$ (blue).}\label{fig:log_illustr}
\end{figure}

To illustrate the R-sens measure in equation~(\ref{eq:Rsenslog}), we simulated 1000 observations from a logistic regression model with
two predictor variables $x_1$ and $x_2$ whose true regression coefficients are $\beta_1 = 1$ and $\beta_2 = 0$.
The predictor variables are independent and normally distributed with zero mean and
standard deviation one.
The R-sens sensitivities for both variables given by equation~(\ref{eq:Rsenslog})
are shown in Figure~\ref{fig:log_illustr}.
The dashed line shows the derivative of the prediction function without the Fisher information term.
Because of the link function, this derivative is not constant and is much larger close to the decision boundary.
The Fisher information term does not remove this effect, but gives a bit more weight to points further away.

\clearpage

\section{ASYMPTOTIC RESULTS FOR GENERALISED LINEAR MODELS}

In Section~2.1 in the main paper, we discussed the behaviour of the posterior
predictive distribution for a Bayesian linear regression model as the number of observations goes to infinity.
In this case, the predictive uncertainty tends to a constant, and
the R-sens local sensitivity measure for each predictor is proportional the absolute value
of the maximum likelihood estimate of the regression coefficient, $|\widehat{\beta}_d|$.
In this section we discuss the asymptotic results of the logistic and Poisson regression models, which
are nonlinear models that can be used to model binary or integer data.

\subsection{Logistic Regression Model}

\subsubsection{Logit Link Function}

The predictive distribution of a logistic regression model is a Bernoulli distribution.
In the asymptotic limit, the posterior of the regression coefficients concentrates to a point $\widehat{\boldsymbol{\beta}}$,
and the ``success probability" parameter as a function of the predictor variables
is the logistic function
\begin{nalign*}
p(y^{*} = 1 | \mathbf{X}, \mathbf{y}, \mathbf{x}^{*}) := \pi^{*} (\mathbf{x}^{*}) = \frac{\text{exp} (\mathbf{x}^{*} \widehat{\boldsymbol{\beta}})}{1 + \text{exp} (\mathbf{x}^{*} \widehat{\boldsymbol{\beta}})} .
\end{nalign*}
The derivative of $\pi^{*}$ with respect to $x^{*}_d$ is
\begin{nalign*}
\frac{\partial \pi^{*}}{\partial x^{*}_d} = \widehat{\beta}_d \, \pi^{*} (1 - \pi^{*}) .
\end{nalign*}
The Fisher information of the Bernoulli distribution
\begin{nalign*}
\mathcal{I}_{\text{bern}} (\pi^{*}) = \frac{1}{\pi^{*} (1 - \pi^{*})} .
\end{nalign*}
In the limit when the number of observations goes to infinity, the R-sens measure thus evaluates to
\begin{nalign*}
\sqrt{ \mathcal{I}_{\text{bern}} (\pi^{*})  \left ( \frac{\partial \pi^{*}}{\partial x^{*}_d}  \right )^2 } = | \widehat{\beta}_d | \, \sqrt{ \pi^{*} (1 - \pi^{*})} .
\end{nalign*}
The R-sens importance measure for the logistic regression model is proportional to the absolute value of the regression coefficient.
In addition, due to the logistic (inverse) link function, the local importance measure
is higher for
points close to the decision boundary $p(y^{*} = 1) = 0.5$ compared to
points further away. Because the term $\sqrt{ \pi^{*} (1 - \pi^{*})}$ is the same for each predictor variable,
ranking the variables with R-sens is equivalent to ranking with the absolute regression coefficients
$| \widehat{\beta}_d |$ in the limit of infinite data.

Contrary to the linear regression example in the main paper, in logistic regression, the
R-sens measure gives more importance to observations further from the decision boundary.
It can be interpreted in the sense that the derivative of the logistic prediction function, $\frac{\partial \pi^{*}}{\partial x^{*}_d} = \widehat{\beta}_d \, \pi^{*} (1 - \pi^{*})$ gives too much emphasis
to points near the decision boundary, and the R-sens measure makes the sensitivity measure
more even.

\subsubsection{Inverse Normal Link Function}

Now, the ``success probability" parameter as a function of the predictor variables
is the cumulative Normal distribution function
\begin{nalign*}
p(y^{*} = 1 | \mathbf{X}, \mathbf{y}, \mathbf{x}^{*}) := \pi^{*} (\mathbf{x}^{*}) = \Phi (\mathbf{x}^{*} \widehat{\boldsymbol{\beta}}) .
\end{nalign*}
The derivative of $\pi^{*}$ with respect to $x^{*}_d$ is
\begin{nalign*}
\frac{\partial \pi^{*}}{\partial x^{*}_d} = \widehat{\beta}_d \, \mathcal{N} (\mathbf{x}^{*} \widehat{\boldsymbol{\beta}}) .
\end{nalign*}
The Fisher information of the Bernoulli distribution
\begin{nalign*}
\mathcal{I}_{\text{bern}} (\pi^{*}) = \frac{1}{\pi^{*} (1 - \pi^{*})} .
\end{nalign*}
In the limit when the number of observations goes to infinity, the R-sens measure thus evaluates to
\begin{nalign*}
\sqrt{ \mathcal{I}_{\text{bern}} (\pi^{*})  \left ( \frac{\partial \pi^{*}}{\partial x^{*}_d}  \right )^2 } = | \widehat{\beta}_d | \, \frac{\mathcal{N} (\mathbf{x}^{*} \widehat{\boldsymbol{\beta}})}{ \sqrt{ \pi^{*} (1 - \pi^{*})}} .
\end{nalign*}

\subsection{Poisson Regression Model}

In a Poisson regression model, the
predictive distribution is a Poisson distribution.
Here, we consider the commonly used logarithmic link function, where
the mean of the predictive distribution is
\begin{nalign*}
\text{E} [y^{*}] = \text{exp} (\mathbf{x}^{*} \boldsymbol{\beta}) .
\end{nalign*}
In the asymptotic limit of infinite data, the posterior of the regression coefficients concentrates to a point $\widehat{\boldsymbol{\beta}}$,
and the mean of the predictive distribution is given by the exponential function
\begin{nalign*}
\text{E} [y^{*}] = \text{exp} (\mathbf{x}^{*} \widehat{\boldsymbol{\beta}}) .
\end{nalign*}
The derivative of $\text{E} [y^{*}]$ with respect to $x^{*}_d$ is
\begin{nalign*}
\frac{\partial \text{E} [y^{*}]}{\partial x^{*}_d} = \widehat{\beta}_d \, \text{exp} (\mathbf{x}^{*} \widehat{\boldsymbol{\beta}}) .
\end{nalign*}
The Fisher information of the Poisson distribution is
\begin{nalign*}
 \frac{1}{\text{E} [y^{*}]} .
\end{nalign*}
In the limit when the number of observations goes to infinity, the R-sens measure thus evaluates to
\begin{nalign*}
| \widehat{\beta}_d | \,  \text{exp} \left ( \frac{1}{2} \mathbf{x}^{*} \widehat{\boldsymbol{\beta}} \right ) .
\end{nalign*}
The R-sens measure depends on $\mathbf{x}^{*}$, but the additional factor is the same for
all predictors.

\end{document}